\documentclass[aps,showpacs,preprintnumbers,amsmath,amssymb,twocolumn,superscriptaddress,floatfix,nofootinbib,10pt]{revtex4-1}

\usepackage[linesnumbered,lined,boxed,commentsnumbered,ruled,vlined]{algorithm2e}
\usepackage{algpseudocode}
\usepackage{amsfonts}
\usepackage{amsmath}
\usepackage{amssymb}
\usepackage{bm,bbm}
\usepackage{booktabs}
\usepackage{caption}
\usepackage{centernot}
\usepackage{colortbl}
\usepackage{comment}
\usepackage{dcolumn}
\usepackage{epstopdf}
\usepackage{float}
\usepackage{graphicx}
\usepackage[bottom]{footmisc}
\usepackage{mathrsfs}
\usepackage{mathtools}
\usepackage{multirow}
\usepackage{physics}
\usepackage{ragged2e}
\usepackage{subfigure}
\usepackage{tabularx}
\usepackage{tikz}
\usepackage{verbatim}
\usepackage{xkcdcolors}
\usepackage{hyperref}
\usepackage{cleveref}

\DeclareCaptionJustification{justified}{\justifying}
\hypersetup{colorlinks=true, citecolor=orange, urlcolor=blue, linkcolor=magenta}

\newtheorem{theorem}{Theorem}

\newtheorem{remark}[theorem]{Remark}

\begin{document}

\title{Global Synchronization in Matrix-Weighted Networks}

\author{Anna Gallo}
\affiliation{IMT School for Advanced Studies, Piazza San Francesco 19, 55100 Lucca (Italy)}
\affiliation{INdAM-GNAMPA Istituto Nazionale di Alta Matematica `Francesco Severi', P.le Aldo Moro 5, 00185 Rome (Italy)}

\author{Yu Tian}%
\affiliation{Center for Systems Biology Dresden, 01307 Dresden, Germany}
\affiliation{Max-Planck Institute for Molecular Cell Biology and Genetics,
01307 Dresden, Germany}
\affiliation{Cluster of Excellence, Physics of Life, TU Dresden, 01307 Dresden, Germany}
\affiliation{Max Planck Institute for the Physics of Complex Systems, 01187 Dresden, Germany}

\author{Renaud Lambiotte}
\email{renaud.lambiotte@maths.ox.ac.uk}
\affiliation{Mathematical Institute, University of Oxford, Woodstock Rd, Oxford OX2 6GG (United Kingdom)}

\author{Timoteo Carletti}
\affiliation{Department of mathematics and Namur Institute for Complex Systems, naXys\\University of Namur, Rue Graf\'e 2, B5000 Namur (Belgium)}

\date{\today}

\begin{abstract}
Synchronization phenomena in complex systems are fundamental to understanding collective behavior across disciplines. While classical approaches model such systems by using scalar-weighted networks and simple diffusive couplings, many real-world interactions are inherently multidimensional and transformative. To address this limitation, Matrix-Weighted Networks (MWNs) have been introduced as a versatile framework where edges are associated with matrix weights that encode both interaction strength and directional transformation. In this work, we investigate the emergence and stability of global synchronization (GS) in MWNs by studying coupled Stuart-Landau (SL) oscillators, an archetypal model of nonlinear dynamics near a Hopf bifurcation. Besides the SL, we considered a generalization of regular oscillators to higher dimensions and also the Lorenz model as a prototype of chaotic oscillators. We derive a generalized Master Stability Function (MSF) tailored to MWNs and establish necessary and sufficient conditions for GS to occur. Central to our analysis is the concept of coherence, a structural property of MWNs ensuring path-independent transformations. Our results show that coherence is necessary to have global synchronization and provides a theoretical foundation for analyzing multidimensional dynamical processes in complex networked systems.
\end{abstract}


\maketitle
\section{Introduction}
\label{sec:intro}

Networks have emerged as a powerful framework for modeling and understanding complex systems across different disciplines such as sociology, economics, biology, and technology~\cite{newman2018networks}. 
The structure of a network plays a crucial role in shaping the behavior of dynamical processes that occur on it~\cite{porter2016dynamical}. In the most straightforward scenario, namely once we consider a linear dynamical system, this interplay is entirely governed by the spectral characteristics of a matrix that represents the graph, typically the adjacency matrix or the graph Laplacian. 
This equivalence allows to identify network features facilitating or hindering spreading processes, but also to design efficient tools to identify key structural aspects of a network, such as identifying influential nodes or detecting communities~\cite{lambiotte2021modularity}. 

The study of non-linear dynamics on networks is a more challenging task, often requiring approximations, e.g., heterogeneous mean-field \cite{moreno2002epidemic}, or techniques designed for specific types of models \cite{chandra2019complexity}, to make analytical progress. 
A key family of dynamic phenomena in complex networks is synchronization, i.e., the spontaneous emergence of collective oscillatory behavior between interacting units~\cite{strogatz2003synchronization,osipov2007synchronization,arenas2008synchronization}. Synchronization is ubiquitous, manifesting in systems as diverse as electric power grids, neural networks, and social coordination~\cite{vallacher2005dynamics,motter2019spontaneous,muller2022neural,klickstein2019symmetry}.

Traditionally, synchronization has been studied within the framework of real scalar-weighted networks, where interactions between nodes are modeled by using simple diffusive couplings~\cite{fujisaka1983stability,Pecora_etal97}. However, in many real-world contexts, nodes can interact through multiple and more complex connectivity patterns: they may involve multidimensional states or act through transformations such as rotations or projections. Such multidimensional interactions have been commonly represented by multilayer networks~\cite{kivela2014multilayer,sorrentino2020group}, where synchronization phenomena have been investigated with particular focus on the interplay between intra-layer topology and inter-layer coupling~\cite{zhang2015explosive,della2020symmetries,leyva2017inter}. Let us also mention the case of Global Topological (Dirac) Synchronization~\cite{carlettiPRL2023,Carletti_2025} where signals of (possibly) different dimensions are defined on the faces of a simplicial complex and dynamically interact to eventually determine global synchronization.

Recent years have seen significant advances in understanding nonlinear dynamics in different settings. For instance, synchronized dynamics have been explored on temporal networks~\cite{ghosh2022synchronized}, and various collective states beyond global synchronization, such as clustering and partial synchrony, have been investigated on weighted networks~\cite{jenifer2024synchronizability,anwar2021relay}. Other complementary studies have explored generalizations of the Stuart-Landau model to higher-dimensional systems~\cite{gogoi2024exactly} and of the Kuramoto model to multidimensional interactions~\cite{buzanello2022matrix,fariello2024exploring}.

More recently, Matrix-Weighted Networks (MWNs) have been introduced as a novel framework to capture the complexities of multidimensional interactions~\cite{tian2025matrix}. In a MWN, each edge is associated with a matrix that encapsulates both the strength and the transformative nature of the connection between nodes. Specifically, this matrix can be decomposed into a magnitude and a directional transformation component, enabling the network to represent how multi-dimensional information flows and is mixed as it passes from a node to its neighbors. Let us observe that the works by~\cite{fujisaka1983stability,Pecora_etal97} can be seen as a very peculiar and simple case of MWN where all the weight matrices are equal to each other and thus independent from the edges.

We note that the MWNs are also closely related to the notion of connection graphs in the mathematics community, defined for different purposes. The corresponding graph connection Laplacian was proposed by Singer and Wu when studying the transformations among massive high-dimensional datasets, where they showed the convergence to the connection Laplacian operator (for vector fields over the manifold) \cite{singer2012image}. Researchers also analyzed the properties of the graph connection Laplacian via the Cheeger inequality \cite{bandeira2013cheeger}, along with its applications in clustering and ranking  \cite{chung2013local,chung2012ranking}. Recently, the connection Laplacian has also been extended for simplicial complexes to effectively incorporate the directionality of simplices \cite{gong2024higher}.  

In this work, we explore the emergence and stability of \textit{Global Synchronization} (GS) in MWNs by investigating the nonlinear dynamics of coupled Stuart-Landau (SL) oscillators, a canonical model that captures the behavior of nonlinear systems near a supercritical Hopf bifurcation~\cite{landau1944problem,stuart1958non,stuart1960non,Stuart1978,vanharten,aranson,garcamorales}. Due to its generality and analytical tractability, such a model is a paradigm for studying the phenomenon of multi-dimensional synchronization in complex systems. Here, we adapt the standard SL oscillators to the setting of MWNs, derive a generalized \textit{Master Stability Function} (MSF) for this framework, and provide necessary and sufficient conditions for the existence and the stability of synchronized solutions. Besides the SL model, we benchmarked our study by considering an example of a higher-dimensional dynamical system exhibiting regular oscillations, and then also the Lorenz system to emphasize the generality of the proposed setting dealing with global synchronization of chaotic oscillators. In this analysis, we explore the role of a key structural property of MWNs, known as \textit{coherence}~\cite{tian2025matrix}. Specifically, a MWN is said to be coherent if the product, i.e., the composition, of the {transformation component of} matrix weights along every oriented cycle equals the identity transformation. Intuitively, this means that any signal propagating through multiple paths in the MWN always returns to its starting point without distortion. 

The emergence of GS is then proved by resorting to linear stability analysis of the system dynamics close to a periodic reference solution. This allows to introduce a Laplace operator whose spectrum exhibits a suitable property to ensure GS; let us emphasize that coherence is a necessary ingredient for this property to hold true. Eventually, we bring to the fore the need for an interesting interplay between the MWN structure, e.g., its matrix weights, and the dynamical system: the latter should preserve the MWN coherence.

By accounting for the structure and directionality of interactions in multiple dimensions, our framework provides a versatile tool to study emergent collective behaviors in complex systems and may inform the design and control of real-world networks. Indeed, MWNs and their synchronization dynamics have potential applications across several real-world contexts. For instance, they can model multidimensional interactions in social networks, capture coordinated dynamics in neural circuits, or describe synchronization phenomena in power grids. Moreover, in neuroscience, interactions between brain regions often involve multidimensional signals (e.g., oscillatory activities), where transformations such as rotations may naturally occur. Finally, in graph learning, and more specifically in graph convolutional networks, node features are diffused and aggregated to capture non-local dependencies in graphs~\cite{di2022understanding}. These examples show how the proposed framework not only extends theoretical synchronization analysis but also provides a natural foundation for studying collective dynamics in multidimensional systems.

This contribution investigates GS in MWNs by developing a general theoretical framework that captures the mechanisms underlying this collective behavior. Starting from the Stuart–Landau model, we derive analytical conditions for the emergence and stability of GS and validate the theory through dedicated numerical simulations. The proposed framework is then extended to multidimensional dynamical systems of arbitrary dimension $d\ge 2$, including a detailed analysis of coupled Lorenz oscillators as a representative case. The study concludes by outlining the broader implications of these findings and discussing potential directions for future research on synchronization phenomena in complex networked systems.

\section{Results}
\label{sec:res}

Let us consider a generic dynamical system whose state variable, $\vec{x}\in\mathbb{R}^d$, evolves according to 
\begin{equation}
\label{eq:dynsys}
\frac{d\vec{x}}{dt}=\vec{f}(\vec{x})\, ,
\end{equation}
for some nonlinear smooth function $\vec{f}:\mathbb{R}^d\rightarrow \mathbb{R}^d$. Depending on the choice of $\vec{f}$, the system of equations may produce regular or even chaotic oscillatory behavior for $\vec{x}(t)$. 

Let us then assume to have $n$ identical copies of the dynamical system~\eqref{eq:dynsys}, each one being identified by the state variable $\vec{x}_i\in\mathbb{R}^d$, $i=1,\dots,n$, and assume moreover to couple them via a {\em Matrix-Weighted Network}~\cite{tian2025matrix} (MWN), where to any existing link connecting nodes $i$ and $j$, we associate the weight matrix $\mathbf{W}_{ij}\in \mathbb{R}^{d\times d}$. By anticipating on the following (see Methods Section~\ref{sec:meth}), we define the scalar weight $w_{ij}:=||\mathbf{W}_{ij}||_2$ and thus rewrite $\mathbf{W}_{ij}=w_{ij}\mathbf{R}_{ij}$, where $w_{ij}$ represents the magnitude and the $d\times d$ matrix $\mathbf{R}_{ij}$ the transformation; in this work the latter will belong to the group of rotations. Starting from the latter matrices, one can define the {\em supra-Laplacian matrix}, $\mathcal{L}$, whose structure recalls the one of combinatorial graph Laplacian, i.e., with ``node degree'' on the diagonal and negated ``adjacency matrix'' off the diagonal (see~\cite{tian2025matrix} and Methods Section~\ref{sec:meth}). The time evolution of the state variable anchored to the $i$--th node, $\vec{x}_i$, is then described by
\begin{align}
\label{eq:dynsyscoupled2}
\frac{d\vec{x}_i}{dt}&= \vec{f}(\vec{x}_i)- \sum_j \mathcal{L}_{ij}\vec{h}(\vec{x}_j)\\
&\, =\vec{f}(\vec{x}_i) -\sum_j {w}_{ij}\mathbf{I}_d \vec{h}(\vec{x}_i) + \sum_j \mathbf{W}_{ij}\vec{h}(\vec{x}_j)\\
&\,=\vec{f}(\vec{x}_i) -\sum_j {w}_{ij}\mathbf{I}_d \vec{h}(\vec{x}_i) + \sum_j w_{ij}\mathbf{R}_{ij}\vec{h}(\vec{x}_j)\, ,
\end{align}
where $w_{ij}$ is the real scalar weight associated to the edge $(i,j)$, $\mathbf{R}_{ij}$ the transformation and we assume that the coupling function $\vec{h}$ does not depend on the node index.

By defining the vector $\vec{x}=(\vec{x}_1^\top,\dots,\vec{x}_n^\top)^\top\in \mathbb{R}^{nd}$, we can rewrite Eq.~\eqref{eq:dynsyscoupled2} as follows
\begin{equation}
\label{eq:dynsyscoupledLmw2}
\frac{d\vec{x}}{dt}={f}_*(\vec{x})- \mathcal{L}h_*(\vec{x})\, ,
\end{equation}
where $f_*$ and $h_*$ act component-wise resulting into a $nd$-dimensional vector, namely
\begin{align}
 {f}_*(\vec{x}):=(\vec{f}(\vec{x}_1)^\top,\dots,\vec{f}(\vec{x}_n)^\top)^\top\, ,
\end{align}
and similarly for $h_*$.

Let us, now, consider a reference solution, $\vec{s}(t)$, of the isolated system~\eqref{eq:dynsys}, then {\em Global Synchronization} (GS) amounts to require $\vec{x}_i(t)\equiv\vec{s}(t)$, for all $i=1,\dots, n$ and $t\geq 0$, to be a stable solution of the interconnected system~\eqref{eq:dynsyscoupled2}. If the underlying support is a connected network with real valued weights, the existence of the eigenvector-eigenvalue pair $\vec{1}_n=(1,\dots,1)^\top\in\mathbb{R}^n$, $\Lambda^{(1)}=0$, ensures that $\vec{x}_i(t)=\vec{s}(t)$ for all $i=1,\dots, n$, is always a solution~\cite{fujisaka1983stability,Pecora_etal97,pecora1998master}. 

We will hereby show that this is not the case for MWNs. 
We will be able to determine a necessary and sufficient condition to ensure the latter to hold true. By anticipating the following, the {\em coherence} condition will reveal an interplay between the topology of the network and the matrix weights. Let us also observe that a similar constraint was recently shown to hold true in the context of global topological synchronization on simplicial complexes, both in the case of coupled topological signals of the same dimension~\cite{carlettiPRL2023} and, by leveraging on the property of the Dirac operator, {when} topological signals of different dimensions do interact~\cite{Carletti_2025}.

{To establish the results presented above, a key role is played by the notion of \textit{coherence} in a MWN, which we formally hereby introduce (see~\cite{tian2025matrix} for further details). A MWN is said to be \textit{coherent} if, for every oriented cycle composed by $k$ different edges, $\mathcal{C}:=((i_1,i_2),(i_2,i_3),\dots,(i_k,i_1))$, the product of the transformation matrices along the cycle equals the identity, i.e., $\prod_{(i,j)\in\mathcal{C}}\mathbf{R}_{ij}=\mathbf{I}_d$. This condition ensures that the net transformation along any cycle is neutral and it will have important consequences concerning the existence of global synchronization.
 This also implies that nodes can be partitioned into distinct groups such that the transformation of any walk between nodes within the same group is the identity. Importantly, the transformation from any node in one group to any node in another group will be the same. In the following, we focus on the case when the transformation on each edge is a rotation, and we denote by $\mathbf{O}$ the matrix associated with a walk, {i.e., the composition of the corresponding sequence of rotations.}
Since the composition of two rotations is itself a rotation, $\mathbf{O}$ is also a rotation (see Fig.~\ref{fig:CohNet1}). 
\begin{figure*}[ht!]
\includegraphics[width=0.9\textwidth]{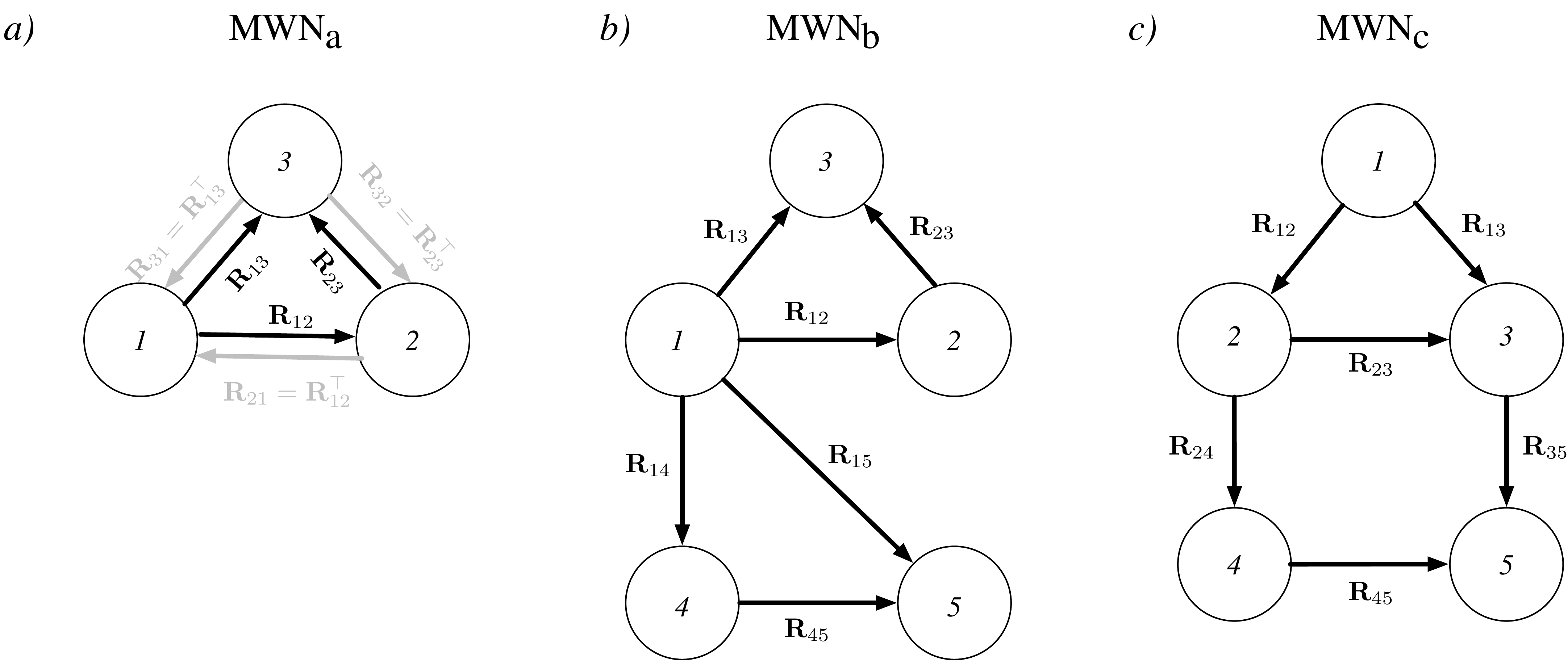}      
\caption{\textbf{Examples of coherent MWNs}. In panel \textit{a)}, we show a triangular network whose matrix weights are denoted by $\mathbf{R}_{ij}$, notice that the matrix associated to each reciprocal link is the transpose of the direct one (shown in light gray). Moreover  $\mathbf{R}_{13}=\mathbf{R}_{12}\mathbf{R}_{23}$ is the condition to have coherence. In panel \textit{b)}, we report a MWN obtained by joining two triangles; for the sake of simplicity, we only show directed weighted matrices, the reciprocal ones are assumed to be given by the transpose as in panel a. Moreover, we assume the following conditions hold true to ensure the coherence: $\mathbf{R}_{13}=\mathbf{R}_{12}\mathbf{R}_{23}$ and $\mathbf{R}_{15}=\mathbf{R}_{14}\mathbf{R}_{45}$. In panel \textit{c)}, we propose a MWN obtained by gluing a triangle and a square; once again, the matrix weights of the reciprocal edges are given by the transpose of the original matrix weight. The coherence condition is given by: $\mathbf{R}_{12}\mathbf{R}_{23}=\mathbf{R}_{13}$ and $\mathbf{R}_{24}\mathbf{R}_{45}=\mathbf{R}_{23}\mathbf{R}_{35}$.}
\label{fig:CohNet1}
\end{figure*}

In a coherent MWN, one can define the block diagonal matrix $\mathcal{S}$, whose $i$-th block is the $d\times d$ matrix, $\mathbf{O}_{1i}$, defined as the product of transformations, i.e., rotations in the present setting, along any oriented walk starting from node $1$ and ending at node $i$. Let us observe that because of the coherence condition, the first node can be arbitrarily chosen (for the sake of definitiveness we hereby choose this node to have the label $1$) and if several paths exist, also the choice of the path is arbitrary and does not change the results (see~\cite{tian2025matrix} and Methods Section~\ref{sec:meth}). Moreover, in the partition associated with the MWN, we also have that $\mathbf{O}_{1i}$ represents the composed transformation between any node in the group of $1$ and any node in the group of $i$. 
{Let us notice that thanks to the above block structure characterizing the coherent MWNs, we can connect the latter with the \textit{identity-transformed MWN}, where, i.e., the transformation of each edge is the identity matrix, $\mathbf{I}_d$, and whose supra-Laplacian matrix reads
\begin{align}
\label{eq:barL}
\bar{\mathcal{L}}=\mathcal{S}\mathcal{L}\mathcal{S}^\top\, ,
\end{align}
(see Methods Section~\ref{sec:meth} for more details). Additionally, by leveraging the definition of the matrix $\mathcal{S}$,} one can then eventually prove that given any $\vec{u}\in\mathbb{R}^{d}$, the vector $\vec{v}=\mathcal{S}^\top(\vec{1}_n\otimes \vec{u})$ is an eigenvector of ${\mathcal{L}}$ with eigenvalue $0$. {As we will show,} the latter vector replaces the synchronous manifold in the case of MWN. 

In Fig.~\ref{fig:CohNet1}, we report some examples of small coherent {MWNs}. Based on the above definition (see also Methods), one can easily show that the matrix $\mathcal{S}$ for the triangle, $\mathrm{MWN_a}$ (panel a of Fig.~\ref{fig:CohNet1}) is given by
\begin{equation}
\label{eq:Sex1}
\mathcal{S}=
\begin{pmatrix}
\mathbf{I}_d & 0 & 0\\
0 & \mathbf{R}_{12} & 0\\
0 & 0 & \mathbf{R}_{13}
\end{pmatrix}\, ,
\end{equation}
namely $\mathbf{O}_{1j}=\mathbf{R}_{1j}$ for $j=2,3$.. Note that in that case, the matrices associated with the two possible paths from node $1$ to node $3$, i.e., $\mathcal{C}_1=(1,3)$ and $\mathcal{C}_2=((1,2),(2,3))$, are the same because $\mathbf{R}_{13}=\mathbf{R}_{12}\mathbf{R}_{23}$, as required by the coherence of the MWN. In the case of the MWN composed of two triangles, $\mathrm{MWN_b}$ (panel b of Fig.~\ref{fig:CohNet1}), we can compute
\begin{equation}
\label{eq:Sex2}
\mathcal{S}=
\begin{pmatrix}
\mathbf{I}_d & 0 & 0 & 0 & 0\\
0 & \mathbf{R}_{12} & 0 & 0 & 0\\
0 & 0 & \mathbf{R}_{13} & 0 & 0\\
0 & 0 & 0 & \mathbf{R}_{14} & 0 \\
0 & 0 & 0 & 0 & \mathbf{R}_{15}
\end{pmatrix}\, .
\end{equation}
Once again, the matrices $\mathbf{O}_{1j}$, $j=2,\dots,5$, equal the rotation matrices $\mathbf{R}_{1j}$ because all nodes can be reached from node $1$ with a single hop. Eventually, for the MWN made of a triangle and a square, $\mathrm{MWN_c}$ (panel of c Fig.~\ref{fig:CohNet1}), we get
\begin{equation}
\label{eq:Sex3}
\mathcal{S}=
\begin{pmatrix}
\mathbf{I}_d & 0 & 0 & 0 & 0\\
0 & \mathbf{R}_{12} & 0 & 0 & 0\\
0 & 0 & \mathbf{R}_{13} & 0 & 0\\
0 & 0 & 0 & \mathbf{R}_{12}\mathbf{R}_{24} & 0 \\
0 & 0 & 0 & 0 & \mathbf{R}_{13}\mathbf{R}_{35}
\end{pmatrix}\, .
\end{equation}
Let us notice that $\mathbf{O}_{14}$ is independent, as it should be because of the coherence condition, from the oriented path used to reach node $4$ from node $1$. Indeed, by using the path $((1,3),(3,5),(5,4))$, we obtain 
\begin{equation*}
\mathbf{R}_{13}\mathbf{R}_{35}\mathbf{R}_{54}=\mathbf{R}_{12}\mathbf{R}_{23}\mathbf{R}_{35}\mathbf{R}_{45}^\top\, ,
\end{equation*}
where we used $\mathbf{R}_{13}=\mathbf{R}_{12}\mathbf{R}_{23}$ and the opposite orientation of the link $(5,4)$ with respect to $(4,5)$, the former being thus associated to the transpose matrix. However $\mathbf{R}_{23}\mathbf{R}_{35}=\mathbf{R}_{24}\mathbf{R}_{45}$, from which we can conclude that
\begin{equation*}
\mathbf{R}_{13}\mathbf{R}_{35}\mathbf{R}_{54}=\mathbf{R}_{12}\mathbf{R}_{24}\mathbf{R}_{45}\mathbf{R}_{45}^\top=\mathbf{R}_{12}\mathbf{R}_{24}\, ,
\end{equation*}
that corresponds to the product of transformations of the oriented walk $((1,2),(2,4))$.

Let us observe that the larger the number of edges shared between different cycles, the larger the number of constraints the rotation matrices should satisfy to have a coherent MWN.

To ensure the existence of GS, we require the nonlinear functions $\vec{f}$ and $\vec{h}$ to be invariant under the transformations $\mathbf{O}_{1i}$ introduced above. Namely $\forall i=1,\dots,n$ and all $\vec{x}\in\mathbb{R}^d$, the following conditions must hold true:
\begin{equation}
\label{eq:condfO}
 \mathbf{O}_{1i}\vec{f}(\mathbf{O}_{1i}^\top \vec{x}) = \vec{f}(\vec{x})\quad\text{and}\quad \mathbf{O}_{1i}\vec{h}(\mathbf{O}_{1i}^\top \vec{x}) = \vec{h}(\vec{x}) \, .
\end{equation}

Based on the above assumption, we will prove in the Methods Section~\ref{sec:meth} that, given any reference solution $\vec{s}(t)$ of Eq.~\eqref{eq:dynsys}, then $\vec{S}(t)= \mathcal{S}^\top (\vec{1}_n\otimes \vec{s}(t))$ is a solution of Eq.~\eqref{eq:dynsyscoupledLmw2}. Note that such an assumption allows us to ensure that the generalized synchronous solution $\vec x_i(t)=\mathbf{O}_{1i}\vec s(t)$ is invariant by the flow~\eqref{eq:dynsyscoupledLmw2}; indeed, the commutation of the nonlinear functions $\vec f$ and $\vec h$ with the transformations $\mathbf{O}_{1i}$ guarantees that trajectories starting on the manifold remain on it, so that $\vec S(t)$ is a solution of the dynamical system. 

Establishing the stability of this solution is the final step needed to prove the existence of GS. To achieve this goal, we will perform a linear stability analysis of the  system~\eqref{eq:dynsyscoupledLmw2} close to the solution $\vec{S}$. The time evolution of the deviation vector $\delta\vec{x}=\vec{x}-\vec{S}$ is given at first order by
\begin{equation}
\label{eq:lin2}
\frac{d}{dt}
 \delta \vec{x}_j=\mathbf{J}_{f}(\mathbf{O}_{1j}^\top\vec{s})
 \delta \vec{x}_j-\sum_\ell \mathcal{L}_{j\ell}\mathbf{J}_{h}(\mathbf{O}_{1\ell}^\top\vec{s})
 \delta \vec{x}_\ell\, ,
\end{equation}
where $\delta\vec{x}=(\delta x_1^\top,\dots,\delta x_n^\top)^\top$ and $\mathbf{J}_{f_*}(\vec{S})$, $\mathbf{J}_{h_*}(\vec{S})$ are respectively the Jacobian of ${f_*}$ and ${h_*}$ evaluated on the solution $\vec{S}$. By using the invariance condition~\eqref{eq:condfO}, we can introduce $\delta\vec{y}_j=\mathbf{O}_{1j}\delta\vec{x}_j$ and obtain
\begin{equation}
\label{eq:lin3}
\frac{d}{dt}
 \delta \vec{y}_j=\mathbf{J}_{f}(\vec{s})
 \delta \vec{y}_j-\sum_\ell \bar{{L}}_{j\ell}\mathbf{J}_{h}(\vec{s})
 \delta \vec{y}_\ell\, ,
\end{equation}
{where $\bar{\mathbf{L}}$ is related to the second supra-Laplace matrix $\bar{\mathcal{L}}$ given by~\eqref{eq:barL} (see also Methods Section~\ref{sec:meth}).}

The latter equation returns a linear, non-autonomous system whose size can be huge, {making the stability analysis of $\delta\vec {y}_j$ challenging.}
Hence, to make some analytical progress, {we exploit the spectral decomposition of $\bar{\mathbf{L}}$, namely, we project the deviations $\delta \vec{y}_j$ onto an orthonormal eigenbasis ${ \bar{\phi}^{(\alpha)}}$ of $\bar{\mathbf{L}}$ associated with the nonnegative eigenvalues $\Lambda^{(\alpha)}$, $\alpha = 1,\dots,n$, and rewrite}
\begin{equation}
\delta \vec{y}_j=\sum_\alpha
\delta \hat{y}_\alpha\bar{\phi}^{(\alpha)}_j\, .
\end{equation}
By inserting the latter into Eq.~\eqref{eq:lin3} and exploiting the orthonormality of the eigenbasis, we get
\begin{equation}
\label{eq:lin5}
\frac{d}{dt}
\delta \hat{y}_\alpha = \left[\mathbf{J}_{f}(\vec{s})-\Lambda^{(\alpha)}\mathbf{J}_{h}(\vec{s})\right]
\delta \hat{y}_\alpha\, ,
\end{equation}
$\forall \alpha=1,\dots,n$. The above equation is the extension of the {\em Master Stability Function} to the Matrix-Weighted Network framework, and allows us to conclude about the stability of the solution $\vec{S}$ based on the eigenvalues $\Lambda^{(\alpha)}$ of the supra-Laplacian matrix. Indeed, let $\lambda(\Lambda^{(\alpha)})$ denote the largest Lyapunov exponent of the $1$-parameter family of matrices $\mathbf{J}_{\alpha}=\mathbf{J}_{f}(\vec{s})-\Lambda^{(\alpha)}\mathbf{J}_{h}(\vec{s})$. Then, if $\lambda(\Lambda^{(\alpha)})<0$ for all $\alpha=1,\dots,n$, we can conclude that $\delta \hat{y}_\alpha\rightarrow 0$ and thus $\vec{S}(t)$ is a stable solution, hence GS is proved. On the other hand, if  there exists {at least one} $\alpha>1$ such that $\lambda(\Lambda^{(\alpha)})>0$, then the system~\eqref{eq:dynsyscoupledLmw2} does not exhibit GS.

To summarize, our analysis relies on the following three assumptions: while the first two ensure the existence of the generalized synchronization solution, the third guarantees its linear stability.
\begin{enumerate}
\item\label{enu:it1} \textit{Coherence of the network}. For any oriented cycle $\mathcal{C}:=((i_1,i_2),(i_2,i_3),\dots,(i_k,i_1))$, the product of the transformations satisfies $\prod_{(i,j)\in\mathcal{C}}\mathbf{R}_{ij}=\mathbf{I}_d$. 
\item\label{enu:it2} \textit{Invariance of dynamics}. The nonlinear functions $\vec f$ and $\vec h$ commute with the transformations $\mathbf O_{ij}$, ensuring that the generalized synchronization manifold $\vec x_i = \mathbf O_{1i}\vec s(t)$ is preserved. 
\item\label{enu:it3} \textit{Spectral stability}. The nonzero eigenvalues of the transformed Laplacian $\bar{\mathcal{L}}$ lie in the stability region determined by the MSF, guaranteeing the linear stability of the manifold. In formulas, the eigenvalues of $\bar{\mathcal{L}}$, $\Lambda^{(\alpha)}$, for any $\alpha=2,\dots,n$, must satisfy the condition $\lambda(\Lambda^{(\alpha)})<0$, i.e., the matrix $\mathbf{J}_{\alpha}$ must be stable for all $\alpha$.
\end{enumerate}

If any of the above conditions is not satisfied, synchronization may not occur. Let us also observe that the change of basis realized via the matrix $\mathcal{S}$ is necessary to observe the GS if the latter is allowed; stated differently, the system could exhibit GS but the choice of the coordinates impedes its visualization.

In the rest of this section, we will provide some applications of the above theoretical framework to a system of Stuart-Landau oscillators coupled via a MWN, to an abstract $d$-dimensional example and to the Lorenz model, also coupled via a MWN.

\subsection{Global synchronization of Stuart-Landau oscillators}
\label{sec:SL}
The first system we consider is the Stuart-Landau (SL) model~\cite{Stuart1978,vanharten,aranson,garcamorales}. This is a paradigmatic model of nonlinear oscillators, often invoked for modeling a wide range of phenomena, resulting to be a normal form for systems close to a supercritical Hopf-bifurcation. 

In Cartesian coordinates, the SL can be defined as
\begin{align}\label{eq:SLmatiso}
\frac{d}{dt}
\begin{pmatrix}
 x\\y
\end{pmatrix} =&
\begin{pmatrix}
 \sigma_{\Re} & -\sigma_{\Im}\\
 \sigma_{\Im} & \sigma_{\Re}
\end{pmatrix}
\begin{pmatrix}
 x\\y
\end{pmatrix}\nonumber\\ &-  (x^2+y^2)
\begin{pmatrix}
 \beta_{\Re} & -\beta_{\Im}\\
 \beta_{\Im} & \beta_{\Re}
\end{pmatrix}
\begin{pmatrix}
 x\\y
\end{pmatrix},
\end{align}
where we introduced the complex model parameters $\sigma=\sigma_{\Re}+i\sigma_{\Im}$ and $\beta=\beta_{\Re}+i\beta_{\Im}$. {One can easily show that, if $\sigma_{\Re}>0$ and $\beta_{\Re}>0$, then the SL admits a stable limit cycle solution given by 
\begin{eqnarray}
    \label{eq:limitcycleSL}
    x_{LC}(t)&=\sqrt{\sigma_{\Re}/\beta_{\Re}}\cos(\omega t) \notag \\
    y_{LC}(t)&=\sqrt{\sigma_{\Re}/\beta_{\Re}}\sin(\omega t)\, ,
\end{eqnarray}
where $\omega =\sigma_{\Im}-\beta_{\Im} \sigma_{\Re}/\beta_{\Re}$. Throughout the following, we assume that this condition on $\sigma_{\Re}$ and $\beta_{\Re}$ is satisfied.}

Let us, now, assume to have $n$ identical copies of the SL model~\eqref{eq:SLmatiso} anchored to the nodes of a MWN and coupled via a diffusive-like nonlinear function. The evolution of the $j$-th oscillator is thus given by
\begin{widetext}
\begin{align}
\label{eq:SLmatnet}
\frac{d}{dt}
\begin{pmatrix}
 x_j\\y_j
\end{pmatrix} &=
\begin{pmatrix}
 \sigma_{\Re} & -\sigma_{\Im}\\
 \sigma_{\Im} & \sigma_{\Re}
\end{pmatrix}
\begin{pmatrix}
 x_j\\y_j
\end{pmatrix} -  (x_j^2+y_j^2)
\begin{pmatrix}
 \beta_{\Re} & -\beta_{\Im}\\
 \beta_{\Im} & \beta_{\Re}
\end{pmatrix}
\begin{pmatrix}
 x_j\\y_j
\end{pmatrix}-\sum_\ell \mathcal{L}_{j\ell}\left[ (x_\ell^2+y_\ell^2)^{\frac{m-1}{2}}
\begin{pmatrix}
 \mu_{\Re} & -\mu_{\Im}\\
 \mu_{\Im} & \mu_{\Re}
\end{pmatrix}
\begin{pmatrix}
 x_\ell\\y_\ell
\end{pmatrix}\right]\nonumber\\
&=:\vec{f}(x_j,y_j)-\sum_\ell \mathcal{L}_{j\ell}\vec{h}(x_\ell,y_\ell)\, ,
\end{align}
\end{widetext}
where $\vec{h}(x_\ell,y_\ell):=(x_\ell^2+y_\ell^2)^{\frac{m-1}{2}}
\begin{pmatrix}
 \mu_{\Re} & -\mu_{\Im}\\
 \mu_{\Im} & \mu_{\Re}
\end{pmatrix}
\begin{pmatrix}
 x_\ell\\y_\ell
\end{pmatrix}$, and $\mu=\mu_{\Re}+i\mu_{\Im}$ is the complex coupling strength.

Given any $2\times 2$ rotation matrix $\mathbf{R}$, the following invariance conditions are clearly satisfied
\begin{equation}
\label{eq:condfR}
\vec{f}(\mathbf{R} \vec{x}) =  \mathbf{R}\vec{f}(\vec{x})\quad\text{ and }\quad \vec{h}(\mathbf{R} \vec{x}) =  \mathbf{R}\vec{h}(\vec{x})\quad\forall \vec{x}\,;
\end{equation}
indeed, it is trivial to verify that 
$\begin{pmatrix}
 \sigma_{\Re} & -\sigma_{\Im}\\
\sigma_{\Im} &  \sigma_{\Re} 
\end{pmatrix}\mathbf{R} =   \mathbf{R}
\begin{pmatrix}
 \sigma_{\Re} & -\sigma_{\Im}\\
\sigma_{\Im} &  \sigma_{\Re} 
\end{pmatrix}$, and similarly for $\beta$ and $\mu$. Then, by defining $(x',y')^\top=\vec{x}'=\mathbf{R}\vec{x}$, it is also clear that $(x')^2+(y')^2=x^2+y^2$.
Hence, if the MWN is coherent, then $\vec{S}(t)=\mathcal{S}^\top (\vec{1}_n\otimes \vec{s}_{LC})$ is a solution of~\eqref{eq:SLmatnet}, where we denoted by $\vec{s}_{LC}=(x_{LC},y_{LC})$ the limit cycle solution of an isolated SL oscillator~\eqref{eq:limitcycleSL}.

To prove the stability of the latter solution, namely the existence of GS, we perform the change of coordinates $\left(
\begin{matrix}
 \delta u_j\\ \delta v_j
\end{matrix}
\right)=\mathbf{O}_{1j}\left(
\begin{matrix}
 \delta x_j\\ \delta y_j
\end{matrix}
\right)$, which allows us to reduce the problem to one involving only scalar weights.
We then linearize the resulting system close to the limit cycle solution and project it onto the Laplace eigenbasis to obtain (see Supplementary Note 1 for more details):
\begin{widetext}
\begin{equation}
\label{eq:rhotheta}
    \frac{d}{dt}\left(
\begin{matrix}
 \delta\hat{u}_\alpha\\ \delta\hat{v}_\alpha
\end{matrix}\right)=\left[ \left(
\begin{matrix}
 -2\sigma_{\Re} & 0\\
 -2\beta_{\Im} \frac{\sigma_{\Re}}{\beta_{\Re}} &0
\end{matrix}
\right)-\left(\frac{\sigma_{\Re}}{\beta_{\Re}}\right)^{\frac{m-1}{2}}\Lambda^{(\alpha)}\left(
\begin{matrix}
 m\mu_{\Re} & -\mu_{\Im}\\
 m\mu_{\Im} & \mu_{\Re}
\end{matrix}
\right)\right]\left(
\begin{matrix}
\delta\hat{u}_\alpha\\ \delta\hat{v}_\alpha
\end{matrix}\right)\, ,
\end{equation}
\end{widetext}
where $\delta\hat{u}_\alpha$ and $\delta\hat{v}_\alpha$ denote the components of the perturbation about the limit cycle in the Laplace eigenbasis. Importantly, their convergence to zero implies GS of the SL coupled oscillators all converging to the limit cycle solution.\\

Finally, let us observe that another advantage of using the SL model is that the resulting Jacobian matrix turns out to be constant, even when evaluated along the limit cycle solution, simplifying thus the stability analysis of~\eqref{eq:rhotheta}.\\

Let us now present numerical results to support the theoretical framework discussed above, focusing on SL oscillators coupled via a MWN. By using the coordinates $(u_j,v_j)^\top=\mathbf{O}_{1j}(x_j,y_j)^\top$, we can define the order parameter
\begin{align}
\label{eq:ordpar}
R(t)=\frac{1}{n} \Big\lvert \sum_{j=1}^n (u_j^2+v_j^2)^{1/2}e^{i \xi_j}\Big\rvert\, ,
\end{align}
where $\tan (\xi_j) =v_j/u_j$. {Such a quantity provides a scalar measure of how well the oscillators are synchronized in both phase and amplitude.} If the system synchronizes, then the amplitudes $(u_j^2+v_j^2)^{1/2}$ will converge to the limit cycle amplitude, i.e., $\sqrt{\sigma_{\Re}/\beta_{\Re}}$, while the phase differences $(\xi_j-\xi_\ell)$ will converge to zero; in conclusion, $R(t)$ will converge to $\sqrt{\sigma_{\Re}/\beta_{\Re}}$. On the other hand, if the system does not synchronize, both amplitudes and phase differences will oscillate in time, and thus $R(t)$ will assume ``small'' values.

In Fig.~\ref{fig:NumResEx1Sync}, we present the numerical simulations obtained by using the $3$-nodes network $\mathrm{MWN_a}$ presented in Fig.~\ref{fig:CohNet1}a. The scalar weights, $w_{ij}$, are randomly drawn from the uniform distribution $\mathrm{U}[0,w_{max}]$, with $w_{max}=0.8$, while the rotation matrices are defined as
\begin{equation}
\label{eq:rotmat3nodes}
\mathbf{R}_{12}=\mathbf{R}_{23}=
\begin{pmatrix}
\cos 2\pi/3 & -\sin 2\pi/3\\
\sin 2\pi/3 & \cos 2\pi/3
\end{pmatrix}\, ,
\end{equation}
and
\begin{equation}
\label{eq:rotmat3nodesb}
\mathbf{R}_{13}=
\begin{pmatrix}
\cos 2\pi/3 & \sin 2\pi/3\\
-\sin 2\pi/3 & \cos 2\pi/3
\end{pmatrix}\, .
\end{equation}
It follows that $\mathbf{R}_{13}=\mathbf{R}_{12}\mathbf{R}_{23}$, and therefore MWNa is a coherent network. As one can observe in panel (a) of Fig.~\ref{fig:NumResEx1NoSync}, the Master Stability Function (red dots) is non-positive (notice that the blue line — plotted as an eye-guide — has been obtained by determining the characteristic roots of~\eqref{eq:rhotheta} by considering $\Lambda^{(\alpha)}$ to be a continuous variable) and, as predicted by the theory, the system synchronizes. This is confirmed by the convergence of the order parameter $R(t)$ (panel (b)) and by the aligned time evolution of the first component of the SL oscillator, $u_j(t)$ (panel (c)).
Let us emphasize the following fact: because the MSF could achieve positive values close to the origin (i.e., the blue curve first increases from $0$ and then reaches negative values), there can be MWNs with suitable Laplace eigenvalues returning a positive MSF and thus preventing the system from synchronization. This is the aim of the next example. 
\begin{figure*}[htbp]        
\includegraphics[width=\textwidth]{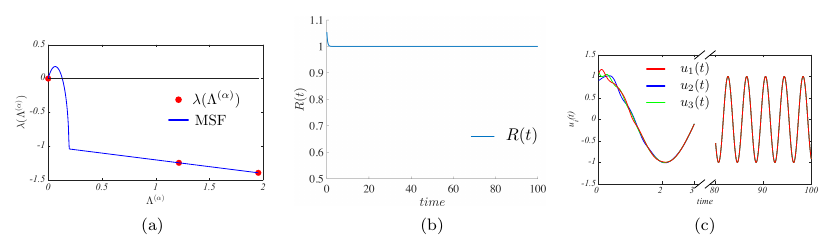}
\caption{\textbf{Global Synchronization of SL coupled via the $\mathrm{MWN_a}$ shown in Fig.~\ref{fig:CohNet1}a}. By visual inspection of the Master Stability Function presented in panel (a), one can clearly appreciate that the latter is negative for all $\Lambda^{(\alpha)}>0$ (red dots), ensuring thus the onset of GS as testified by the order parameters reaching the constant value $\sqrt{\sigma_{\Re}/\beta_{\Re}}=1$ (panel (b)) and the time evolution of $u_j(t)$ for short and long times (panel (c)). The remaining model parameters are given by $\sigma = 1.0 - i 0.5$, $\beta = 1.0+i 1.1$, $m = 3$ and $\mu = 0.1-i 5.5$. The scalar weights have been drawn from $\mathrm{U}[0,0.8]$ and the rotation matrices have been set to $
\mathbf{R}_{12}=\mathbf{R}_{23}=
\begin{pmatrix}
\cos 2\pi/3 & -\sin 2\pi/3\\
\sin 2\pi/3 & \cos 2\pi/3
\end{pmatrix}$ and $
\mathbf{R}_{13}=
\begin{pmatrix}
\cos 2\pi/3 & \sin 2\pi/3\\
-\sin 2\pi/3 & \cos 2\pi/3
\end{pmatrix}$.}
 \label{fig:NumResEx1Sync}
\end{figure*}

GS can be destroyed by changing the scalar weights, while keeping the rest unchanged. In Fig.~\ref{fig:NumResEx1NoSync}, we report numerical simulations to prove this claim; they have been obtained with the same network and parameters used in Fig.~\ref{fig:NumResEx1NoSync} except for the scalar weights, that, now, are drawn from the uniform distribution $\mathrm{U}[0,w_{max}]$, where $w_{max}=0.1$. This rescaling of the weights is expected to shift some eigenvalues into the unstable region. In this case, indeed, one can observe that there exists one $\Lambda^{(\alpha)}$ for which the Master Stability Function is positive (panel (a)). As a result, the order parameter $R(t)$ now oscillates about $0.2$ (panel (b)), and some nontrivial patterns for the variable $u_j(t)$ emerge (panel (c)).

\begin{figure*}[htbp]
        \includegraphics[width=\textwidth]{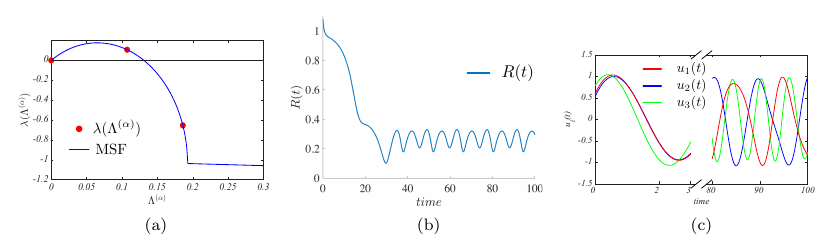}
        \caption{\textbf{Absence of Global Synchronization of SL coupled via the $\mathrm{MWN_a}$ shown in Fig.~\ref{fig:CohNet1}a}. The Master Stability Function (panel (a)) assumes positive values, hence the order parameter (panel (b)) cannot converge to $\sqrt{\sigma_{\Re}/\beta_{\Re}}=1$ and $u_j(t)$ oscillates out of phase for short and long times (panel (c)). The remaining model parameters and rotation matrices have been set to the same values of Fig.~\ref{fig:NumResEx1Sync} but $w_{ij}$, now drawn from $\mathrm{U}[0,0.1]$.}
        \label{fig:NumResEx1NoSync}
\end{figure*}

To study the impact of the coherence condition and the lack thereof, we consider SL oscillators coupled via the $\mathrm{MWN_b}$ shown in Fig.~\ref{fig:CohNet1}b. 
For the ``upper'' triangle of the network, let us fix the rotation matrices as
\begin{equation}
\label{eq:O1jb}
\mathbf{R}_{12}=\mathbf{R}_{23}=
\begin{pmatrix}
\cos \theta_1 & -\sin\theta_1\\
\sin\theta_1 & \cos \theta_1
\end{pmatrix}\, ,
\end{equation}
and
\begin{equation}
\label{eq:O1jb2}
\mathbf{R}_{13}=
\begin{pmatrix}
\cos 2\theta_1 & -\sin 2\theta_1\\
\sin 2\theta_1 & \cos 2\theta_1
\end{pmatrix}\, ,
\end{equation}
so that for all fixed $\theta_1\in [0,2\pi)$ the relation $\mathbf{R}_{13}=\mathbf{R}_{12}\mathbf{R}_{23}$ holds true. 
Similarly, for the ``lower'' triangle, we define 
\begin{equation}
\label{eq:O1jb3}
\mathbf{R}_{14}=\mathbf{R}_{45}=
\begin{pmatrix}
\cos \theta_2 & -\sin\theta_2\\
\sin\theta_2 & \cos \theta_2
\end{pmatrix}\, ,
\end{equation}
for some $\theta_2\in [0,2\pi)$. It then follows that the rotation matrix
\begin{equation}
\label{eq:O1jb4}
\mathbf{R}_{15}=
\begin{pmatrix}
\cos \theta & -\sin \theta\\
\sin \theta & \cos \theta
\end{pmatrix},
\end{equation}
fulfills the coherence condition $\mathbf{R}_{15}=\mathbf{R}_{14}\mathbf{R}_{45}$ if and only if $2\theta = \theta_2$.

Numerical results are reported in Fig.~\ref{fig:NumResEx2Sync}. In panels (a), (b) and (c), we set $\theta_1=2\pi/3$, $\theta_2=2\pi /5$ and $\theta=\pi/5$, which ensure that WMNb is coherent; while, the scalar weights are now drawn from $\mathrm{U}[0,w_{max}]$, with $w_{max}=0.8$. In this setting, we conclude that the Master Stability Function (panel (a)) is non-positive, and the system synchronizes, as evidenced by the behavior of the order parameter $R(t)$ (panel (b)) and the time evolution of $u_j(t)$ (panel (c)). On the other hand, panels (d), (e) and (f) show the numerical results obtained with the choice $\theta=2\pi/5$ that removes the coherence assumption. Although the Master Stability Function is non-positive (panel (d)), the system does not synchronize: the order parameter converges to a quite small value (panel (e)), and the oscillators $4$ and $5$ are now out of phase (panel (f)). This outcome highlights that the vector $\vec{S} = \mathcal{S}^\top (\vec{1}_5 \otimes \vec{s})$ is no longer a solution to the system, and therefore the Master Stability Function is no longer a valid predictor of global synchronization.

\begin{figure*}[ht!]
 \includegraphics[width=\textwidth]{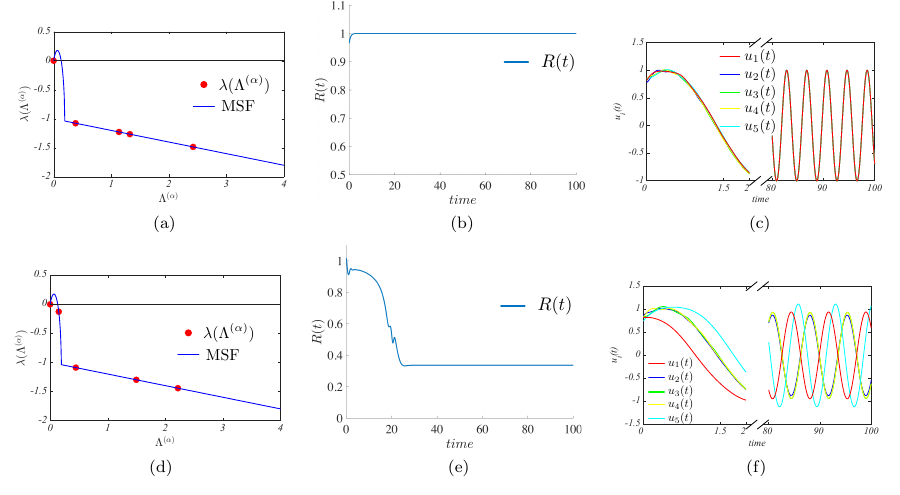}
        \caption{\textbf{Impact of the coherence assumption on GS of SL coupled via the $\mathrm{MWN_b}$ shown in Fig.~\ref{fig:CohNet1}b}. Panels (a), (b) and (c) report simulations obtained under the assumption of coherent network, i.e., we set $\theta_1=2\pi/3$, $\theta_2=2\pi/5$ and $\theta=\pi/5$ in Eqs.~\eqref{eq:O1jb}-~\eqref{eq:O1jb4}; the Master Stability Function (panel (a)) is non-positive and the system converges to synchronization, $R(t)$ stabilizes at $\sqrt{\sigma_{\Re}/\beta_{\Re}}=1$ (panel (b)) and the oscillators are in phase after a transient time (panel (c)). In the panels (d), (e) and (f), the coherence condition does not hold true because we set $\theta=2\pi/5$ in Eq.~\eqref{eq:O1jb4},
        and the system is not capable to synchronize, the order parameter converges to a small value (panel (e)) and two oscillators are out of phase (panel (f)). The model parameters are given by $\sigma = 1.0 - i0.5$, $\beta = 1.0+i1.1$, $m = 3$, $\mu = 0.1-i5.5$ and the scalar weights have been drawn from $\mathrm{U}[0,0.8]$.}
        \label{fig:NumResEx2Sync}
\end{figure*}

\subsection{Global synchronization of higher-dimensional dynamical systems}
\label{sec:rot}

The aim of this section is to present some results within the framework of higher-dimensional dynamical systems, $d>2$, thereby extending the analysis done for the $2$-dimensional Stuart-Landau model.

Before introducing the general case $d>2$, let us consider in detail the $d=3$ one, which contains all the main ideas without adding unnecessary complications. Let us thus assume the isolated system~\eqref{eq:dynsys} to have the following form
\begin{equation}
\label{eq:3Dmodisol}
\frac{d{\vec{x}}}{dt}=\vec{f}(\vec{x}):=\mathbf{A}\vec{x}-|\vec{x}|^2\mathbf{B}\vec{x}\, ,
\end{equation}
where $\vec{x}\in\mathbb{R}^3$ and $\mathbf{A}$ and $\mathbf{B}$ are $3\times 3$ matrices. The specific functional form of the latter system has been chosen to ensure that the system preserves the coherence of the MWN.

In $3-$dimension, rotation matrices can be defined by the Rodrigues formula
\begin{equation}
\label{eq:rodrigues}
    \mathbf R(\vec{u}, \theta) = \mathbf{I}_3 + \sin(\theta) [\vec{u}]_\times + (1 - \cos(\theta)) ([\vec{u}]_\times)^2\, ,
\end{equation}
where the vector $\vec u = (u_x,u_y,u_z)^\top\in\mathbb R^3$ denotes the rotation axis, $\theta$ is the rotation angle in the plane orthogonal to $\vec u$, $\mathbf{I}_3$ is the $3\times 3$ identity matrix, and $[\vec{u}]_\times$ is the skew-symmetric matrix associated to the cross-product of $\vec{u}$, i.e.,
\begin{align}
[\vec{u}]_\times = \begin{pmatrix}
0 & -u_z & u_y \\
u_z & 0 & -u_x \\
-u_y & u_x & 0
\end{pmatrix}\, .
\end{align}
For the sake of definitiveness and without loss of generality, since we can always perform a rotation to map any rotation axis into the vector $(0,0,1)^\top$, we fix $\vec{u}=(0,0,1)^\top$ (see the Supplementary Notes for more details). Then, to ensure the invariance of~\eqref{eq:3Dmodisol} with respect to such rotations, the matrices $\mathbf{A}$ and $\mathbf{B}$ must be of the form:
\begin{align}
\label{eq:ABmat}
\mathbf{A}=
\begin{pmatrix}
\mathbf{A}_1 & \vec{0}\\
\vec{0}^\top & \mu_A 
\end{pmatrix}\quad\text{and}\quad\mathbf{B}=
\begin{pmatrix}
\mathbf{B}_1 & \vec{0}\\
\vec{0}^\top &\mu_B 
\end{pmatrix},
\end{align}
where $\mu_A, \mu_B\in\mathbb R$, $\vec{0}=(0,0)^\top$ and $\mathbf A_1$, $\mathbf B_1$ are $2\times 2$ rotation matrices defined as
\begin{align}
\label{eq:A1mat}
\mathbf{A}_1=\lambda_A
\begin{pmatrix}
\cos(a) & -\sin(a)\\
\sin(a) & \cos(a)
\end{pmatrix},
\end{align}
and
\begin{align}
\label{eq:B1mat}
\mathbf{B_1}=\lambda_B
\begin{pmatrix}
\cos(b) & -\sin(b)\\
\sin(b) & \cos(b)
\end{pmatrix},
\end{align}
with $\lambda_A, \lambda_B\in\mathbb R$. A straightforward computation allows to rewrite $\mathbf{A}$ and $\mathbf{B}$ as
\begin{align}
\label{eq:ABmat2}
\mathbf{A}=
\begin{pmatrix}
\mathbf{A}_1 & \vec{0}\\
\vec{0}^\top & 0 
\end{pmatrix}+\mu_A \mathbf{P}_u\text{ and }\mathbf{B}=
\begin{pmatrix}
\mathbf{B}_1 & \vec{0}\\
\vec{0}^\top &0 
\end{pmatrix}
+\mu_B \mathbf{P}_u\, ,
\end{align}
where $\mathbf{P}_u=\vec{u}\vec{u}^\top$ is the {orthogonal} projection onto the axis $\vec{u}=(0,0,1)^\top$. This corresponds to choose matrices that commute with rotations about the $\vec{u}$-direction, namely, $\mathbf{A}_1$ and $\mathbf{B}_1$ act as rotation (possibly as a contraction or an expansion according to the parameters $\lambda_A$ and $\lambda_B$) in the $xy$-plane.

To study the dynamics of the system~\eqref{eq:3Dmodisol}, we decompose $\vec{x}=(x_1,x_2,x_3)^\top$ as follows:
\begin{equation}
\label{eq:breakdyn3d}
    \vec x = \vec x_{||}+\vec x_{\perp}\, ,
\end{equation}
where $\vec x_{||}=x_{||}\vec u=(\vec u\cdot\vec x)\vec u$ is the component along $\vec u$ and $\vec x_{\perp}=\vec x-\vec x_{||}$ is the one perpendicular to it. In particular, since $\vec u=(0,0,1)^\top$, the decomposition can be explicitly rewritten as
\begin{equation}
\label{eq:xsplit}
\vec{x}=\vec{x}^{||}+\vec x^{\perp}\equiv 
\begin{pmatrix}
\vec{\xi} \\0
\end{pmatrix}+
\begin{pmatrix}
\vec{0}\\x_3
\end{pmatrix}\, ,
\end{equation}
where $\vec{\xi}\in\mathbb{R}^2$. Because of this decomposition, system~\eqref{eq:3Dmodisol} can be rewritten as
\begin{equation}
\label{eq:3Dmodisolbis}
\begin{cases}
 \frac{d{\vec{\xi}}}{dt} &=\mathbf{A}_1\vec{\xi}-\left(|\vec{\xi}|^2+x_3^2\right) \mathbf{B}_1\vec{\xi},\\
 \frac{d{x}_3}{dt}&=\mu_A x_3-\left(|\vec{\xi}|^2+x_3^2\right)\mu_B x_3\, ,
\end{cases}
\end{equation}

To make a step further, we express $\vec{\xi}$ in polar coordinates, i.e., $\xi_1=r\cos\phi$ and $\xi_2=r\sin\phi$, and show (see Supplementary Note 2) {that it admits a stable limit cycle solution given by $\hat{r}(t) = \sqrt{\frac{\lambda_A \cos(a)}{\lambda_B \cos(b)}}$, $\hat{x}_3(t) = 0$, and $\hat{\phi}(t) = \phi(0) + \omega t$, where $\omega = \lambda_A \frac{\sin(a - b)}{\cos(b)}$, provided the following conditions hold:
\begin{equation}
\label{eq:stacond3D1}
\lambda_A \cos(a) > 0,\quad \lambda_B \cos(b) > 0,
\end{equation}
and
\begin{equation}
\label{eq:stacond3D2}
\mu_A \lambda_B \cos(b) - \mu_B \lambda_A \cos(a) < 0.
\end{equation}
(see the Supplementary Note 2 for a complete analysis of the equilibria of~\eqref{eq:3Dmodisolbis} and their stability).}

{Such a limit cycle will serve as the reference solution characterizing GS}
\begin{equation}
    \label{eq:s3Drot}
    \vec{s}(t)=\left(\hat{r}\cos(\omega  t),\hat{r}\sin(\omega t),0\right)^\top\, .
\end{equation}

Let us, now, consider $n$ identical copies of the isolated system~\eqref{eq:3Dmodisol} coupled via a MWN and study the emergence of GS. Specifically, we are interested in studying
\begin{eqnarray}
\label{eq:3Dmodnet}
\frac{d{\vec{x}}_i}{dt}&=&\vec{f}(\vec{x}_i)-\epsilon \mathbf{C} \sum_j \mathcal{L}_{ij}\vec{x}_j\nonumber\\
&=&\mathbf{A}\vec{x}_i-|\vec{x}_i|^2\mathbf{B}\vec{x}_i-\epsilon \mathbf{C} \sum_j \mathcal{L}_{ij}\vec{x}_j\, ,
\end{eqnarray}
$\forall i\in\{1,\dots,n\}$, where $\vec{x}_i\in\mathbb{R}^3$, $\mathbf{A}$ and $\mathbf{B}$ are $3\times 3$ matrices defined as in~\eqref{eq:ABmat2}, and $\mathbf{C}$ is a third $3\times 3$ matrix given by
\begin{align}
\label{eq:Cmat}
\mathbf{C}=
\begin{pmatrix}
\mathbf{C}_1 & \vec{0}\\
\vec{0}^\top & \mu_C 
\end{pmatrix}\quad \text{with} \quad \mathbf{C}_1=\lambda_C\begin{pmatrix}
\cos(c) & -\sin(c)\\
\sin(c) & \cos(c)
\end{pmatrix}\, ,
\end{align}
where $\lambda_C$, $\mu_C\in\mathbb R$. Let us observe that for the sake of pedagogy, we consider a linear coupling in Eq.~\eqref{eq:3Dmodnet}.

Assuming that the network is coherent and the matrix weights are rotations about the axis $\vec{u}=(0,0,1)^\top$, it follows by construction that
\begin{align}
 \mathbf{O}_{1i}\vec{f}(\mathbf{O}_{1i}^\top \vec{s}) = \vec{f}(\vec{s})\quad \forall i=1,\dots,n\,,
\end{align}
where the definition of the matrices $\mathbf{O}_{1i}$ comes from the coherence condition, while $\vec{s}$ is given by~\eqref{eq:s3Drot}.

By defining the vector $\vec{x}=(\vec{x}_1^\top,\dots,\vec{x}_n^\top)^\top\in \mathbb{R}^{3n}$, we can rewrite Eq.~\eqref{eq:3Dmodnet} as follows
\begin{align}
\label{eq:dynsyscoupledLmw3}
\frac{d\vec{x}}{dt}={f}_*(\vec{x})- \epsilon (\mathbf{I}_n\otimes \mathbf{C})\mathcal{L}\vec{x}\equiv {f}_*(\vec{x})- \epsilon \mathcal{C}\mathcal{L}\vec{x}\, ,
\end{align}
where $\mathcal{C}=\mathbf{I}_n\otimes \mathbf{C}$ is the Kronecker product of the two matrices, and $f_*$ acts component-wise resulting into a $3n-$dimensional vector. 

We have thus formulated the problem within the framework of MWN presented earlier. To make some further analytical progress, we left-multiply Eq.~\eqref{eq:dynsyscoupledLmw3} by the block-diagonal matrix $\mathcal{S}$, {and introduce the transformed variable $\vec{y}=\mathcal{S}\vec{x}=(\vec{y}_1,\dots,\vec{y}_n)^\top$. For all $i=1,\dots,n$, this yields}
\begin{align}
\label{eq:3Dmodnety}
\frac{d{\vec{y}}_i}{dt}=\mathbf{A}\vec{y}_i-|\vec{y}_i|^2\mathbf{B}\vec{y}_i-\epsilon \mathbf{C} \sum_j \bar{{L}}_{ij}\vec{y}_j\, ,
\end{align}
where we used the commutativity property of the matrices $\mathbf{A}$, $\mathbf{B}$, $\mathbf{C}$ with (product of) rotation matrices and the definition of the transformed supra-Laplacian $\bar{\mathcal{L}}$.

Global synchronization to the reference solution $\vec{s}(t)$, given by~\eqref{eq:s3Drot}, can be achieved if $\vec{y}_i(t)=\vec{s}(t)$, for all $i=1,\dots,n$, is a stable solution of Eq.~\eqref{eq:3Dmodnety}. To prove this claim, {we analyze the linear stability of the synchronized state by introducing small perturbations. Specifically, we define
\begin{align}
\label{eq:deltapert}
    \vec{y}_j(t)=\vec{s}(t)+
    \begin{pmatrix}
     \hat{r}\cos(\omega t)\delta u_j -  \hat{r}\sin(\omega t)\delta v_j\\
     \hat{r}\sin(\omega t)\delta u_j +  \hat{r}\cos(\omega t)\delta v_j\\
     \delta\eta_j
    \end{pmatrix}\, ,
\end{align}}
where $\delta u_j$, $\delta v_j$ and $\delta\eta_j$ are small functions whose time evolution, once projected into the eigenbasis formed by the eigenvectors of $\bar{\mathbf{L}}$ is determined by (see Supplementary Note 2)
\begin{align}
\label{eq:3Dmodnetyquat}
\begin{dcases}
 \frac{d\delta{u}_{\alpha}}{dt}  = -2\lambda_A \cos(a) \delta u_\alpha\\
 \hspace{2.8cm} -\epsilon \lambda_C \Lambda^{(\alpha)}\left( \cos(c) \delta u_\alpha-\sin(c) \delta v_\alpha\right)\\
 \frac{d\delta{v}_{\alpha}}{dt}  = 
 -2\lambda_A \frac{\cos(a)\sin(b)}{\cos(b)}\delta u_\alpha\\
 \hspace{2.8cm}
 -\epsilon \lambda_C \Lambda^{(\alpha)}\left( \sin(c) \delta u_\alpha+\cos(c) \delta v_\alpha\right) \\
 \frac{d\delta{\eta}_{\alpha}}{dt}  =\left(\mu_A -\frac{\lambda_A \cos(a)}{\lambda_B \cos(b)}\mu_B\right)\delta\eta_\alpha-\epsilon \mu_C \Lambda^{(\alpha)}\delta\eta_{\alpha}\, ,
\end{dcases}
\end{align}
Let us observe that because of the form of the system~\eqref{eq:3Dmodnety} and of the perturbation we chose~\eqref{eq:deltapert}, the linearized system turns out to be time independent and thus a complete study of its stability can be performed (see Supplementary Note 2 for a comprehensive analysis)

In Fig.~\ref{fig:NumResEx1SyncBis}, we report the numerical simulations of the $3$D-system~\eqref{eq:3Dmodnety} defined on the $3$-nodes network $\mathrm{MWN_a}$ shown in Fig.~\ref{fig:CohNet1}a. The rotation matrices are given by Eqs.~\eqref{eq:rotmat3nodes} and~\eqref{eq:rotmat3nodesb} and {all scalar weights are unitary, i.e., $w_{ij}=1$.}
As shown in panel (a), the Master Stability Function (red dots) is non-positive (the blue line is shown for visual reference, and it has been obtained by replacing $\Lambda^{(\alpha)}$ with a continuous variable), and the system synchronizes. Panel (b) confirms this by showing the time evolution of the first component of $\vec{y}_j$, i.e., $\xi_{1,j}(t)$, while panel (c) illustrates the convergence of the third component, $y_{3,i}(t)$, to zero, reflecting the system stabilization along the rotation axis.
\begin{figure*}[ht!]
    \centering
    \begin{tabular}{c}
        \includegraphics[width=\textwidth]{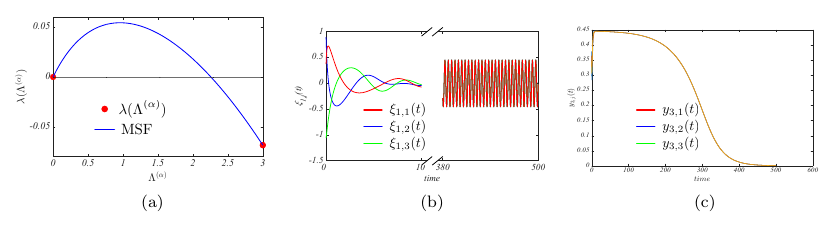}
    \end{tabular}
        \caption{\textbf{Synchronization of the $3$D-system~\eqref{eq:3Dmodnety} on the $\mathrm{MWN_a}$ shown in Fig.~\ref{fig:CohNet1}a with rotation matrices given by~\eqref{eq:rotmat3nodes} and~\eqref{eq:rotmat3nodesb} and scalar weights $w_{ij}=1$}. Panel (a): the Master Stability Function; panel (b): $\xi_{1,j}(t)$, i.e., the first component of $\vec{y}_j$, and panel (c): $y_{3,j}(t)$, for $j=1,2,3$, for short and long times. The matrices, $\mathbf{A}$, $\mathbf{B}$ and $\mathbf{C}$ are defined with $a = 2\pi/5$, $b = 2\pi/3$, $c = 2\pi/7$, $\lambda_A = 1$, $\lambda_B = -3$, $\lambda_C = 2$, $\mu_A = 1$, $\mu_B = 5$ and $\mu_C = 1$. The coupling parameters is fixed to $\epsilon = 0.1$.}
        \label{fig:NumResEx1SyncBis}
\end{figure*}
In Fig.~\ref{fig:NumResEx1NoSyncBis}, we show a set of parameters for which there is no GS. We can observe that the MSF is positive (panel (a)) and the coordinate $\xi_{1,j}$ (panel (b)) clearly shows the absence of synchronization. Additionally, dynamics along the rotation direction exhibit non-trivial behavior (panel (c)).

\begin{figure*}[ht!]
    \centering
    \begin{tabular}{c}
        \includegraphics[width=\textwidth]{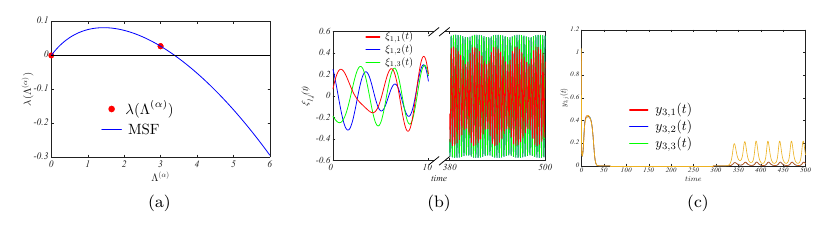}
    \end{tabular}
        \caption{\textbf{Absence of Synchronization for the $3$D-system~\eqref{eq:3Dmodnety} defined on the $\mathrm{MWN_a}$ shown in Fig.~\ref{fig:CohNet1}a with rotation matrices given by~\eqref{eq:rotmat3nodes} and~\eqref{eq:rotmat3nodesb} and scalar weights $w_{ij}=1$}. Panel (a): the Master Stability Function, panel (b): $\xi_{1,j}(t)$, i.e., the first component of $\vec{y}_j$, and panel (c) $y_{3,j}(t)$, for $j=1,2,3$, for short and long times. The matrices, $\mathbf{A}$, $\mathbf{B}$ and $\mathbf{C}$ are defined with $a = 2\pi/5$, $b = 2\pi/3$, $c = 2\pi/7$, $\lambda_A = 1.5$, $\lambda_B = -3$, $\lambda_C = 2$, $\mu_A = 1$, $\mu_B = 5$ and $\mu_C = 1$. The coupling parameters is fixed to $\epsilon = 0.1$.}
        \label{fig:NumResEx1NoSyncBis}
\end{figure*}

\begin{remark} - \textbf{\emph{Higher dimensional dynamical systems, $d>3$.}} 
\emph{The theory presented above can be generalized to $d$-dimensional dynamical systems, $d>3$, coupled via a coherent MWN, provided the vector field preserves the rotation matrices of the MWN.}

\emph{Let us observe that among the $d$-dimensional rotations, one can consider matrices acting on orthogonal $2$-dimensional planes, i.e., the matrix can be written as $2\times 2$ blocks on the diagonal. The function $\vec{f}(\vec{x})$ should thus preserve each $2$-plane, and the analysis reduces to the one presented above, but considering several planes.}

\emph{On the other hand, one can consider the opposite case, where the rotation matrix leaves invariant the direction given by the vector $(0,...,0,1)^\top$ and thus non-trivially acts on $\mathbb{R}^{d-1}$. In this case, we require the function $\vec{f}(\vec{x})$ to preserve this space.}

\emph{Eventually, one can consider combinations of both cases and deal with suitable functions $\vec{f}(\vec{x})$.}
\end{remark}

\subsection{Global synchronization of Lorenz oscillators}
\label{sec:lorenz}
In the previous sections, we have considered regular oscillators, the Stuart-Landau and a generalization in higher dimensions, coupled via a MWN, and shown that global synchronization can be achieved under suitable conditions. The aim of this section is to make a step forward and to consider the case of global synchronization of chaotic oscillators still coupled via a MWN.

For the sake of definitiveness and because it has been largely used in synchronization problems both on networks~\cite{huangpecora2009PRE,sorrentino2024NatComm,boccaletti2024NatComm} and simplicial complexes~\cite{boccaletti2021NatComm}, we decided to consider the Lorenz system~\cite{lorenz1963} as a prototype of chaotic systems. The time evolution of the $i$-th oscillator is thus described by
\begin{align}\label{eq:Lorenzode}
\begin{dcases}
\frac{dx_i}{dt} = \sigma (y_i-x_i)\\
\frac{dy_i}{dt} = x_i (\rho-z_i)-y_i\\
\frac{dz_i}{dt} = x_iy_i-\beta z_i\, ,
\end{dcases}
\end{align}
where $\sigma$, $\rho$ and $\beta$ are model parameters that we hereby fix to $\sigma=10$, $\rho=28$ and $\beta=2$, ensuring thus the existence of a chaotic behavior~\cite{lorenz1963}.

Let us now assume to couple $n$ identical copies of the Lorenz system via a MWN so to obtain
\begin{equation}
    \label{eq:lorenzMWN}
    \frac{d\vec{x}_i}{dt}=\vec{f}(\vec{x}_i)-\epsilon \sum_j \mathcal{L}_{ij}\mathbf{E}\vec{x}_j\, ,
\end{equation}
where $\epsilon>0$ is the coupling strength and we defined
\begin{equation}
\label{eq:Lorenzf}
    \vec{x}_i = \left(\begin{matrix}
        x_i\\y_i\\z_i
    \end{matrix}\right)\, , \, \vec{f}(\vec{x}_i)=\left(\begin{matrix}
        \sigma(y_i-x_i)\\x_i(\rho-z_i)-y_i\\x_iy_i-\beta z_i
    \end{matrix}\right)
\end{equation}
and $\mathbf{E}$ is a constant $3\times 3$ matrix determining the linear coupling. More precisely we assume~\cite{huangpecora2009PRE} all the entries of $\mathbf{E}$ to be zero except one. For instance if $E_{11}=1$, then for all $i\in\{1,\dots,n\}$, $x_i$ (the variable number $1$) will be coupled to $x_j$ (again the variable number $1$) for all nodes $j$ connected to node $i$; if $E_{12}=1$ then $x_i$ will be coupled to $y_j$ (the variable number $2$), and so on.

Let the reference orbit, $\vec{s}(t)$, be the chaotic solution of a single Lorenz oscillator and let us study the conditions under which all nodes do satisfy $\vec{x}_i(t)=\vec{s}(t)$ for all $t\geq 0$. As previously stated, we require the MWN to be coherent, the dynamical system to preserve the coherence, and the spectrum of the Laplace matrix $\bar{\mathcal{L}}$ to return a negative master stability function. For the sake of pedagogy, let us consider a simple MWN composed by four nodes and four links organized to form a square (see panel (c) in Fig.~\ref{fig:LorenzSynch}), moreover, we assume all the rotation matrices to be given by 
\begin{equation*}
\mathbf{R}_\pi=\left(\begin{matrix}
    -1 & 0 & 0\\
    0 & -1 & 0\\
    0 & 0 & 1
\end{matrix}\right)\, ,
\end{equation*}
encoding thus a rotation in the $xy$-plane of an angle $\pi$ about the $z$-axis. Or stating equivalently we have $\mathbf{R}_\pi\cdot (x,y,z)^\top=(-x,-y,z)^\top$. The resulting MWN is trivially coherent because $\mathbf{R}_\pi\mathbf{R}_\pi=\mathbf{I}_3$. On the other hand, the Lorenz system~\eqref{eq:Lorenzode} is clearly invariant with respect to the transformation $(x,y,z)^\top \rightarrow (-x,-y,z)^\top$, namely the function $\vec{f}$ defined in~\eqref{eq:Lorenzf} does satisfy $\mathbf{R}_\pi\vec{f}(\mathbf{R}^\top_\pi\vec{x})=\vec{f}(\vec{x})$, for all $\vec{x}$. Hence, the latter function preserves the coherence of the MWN. As trivially does the linear coupling $\mathbf{E}\vec{x}$.

It remains to prove the negativeness of the MSF. Let us observe that in this case we do not have an analytical expression for the reference solution $\vec{s}(t)=(s_x(t),s_y(t),s_z(t))^\top$ and thus we have to resort to a numerical integration of the linear system~\eqref{eq:lin5} that now reads
\begin{equation*}
\frac{d}{dt}
\delta \hat{y}_\alpha = \left[\mathbf{J}_{f}(\vec{s})-\epsilon \Lambda^{(\alpha)}\mathbf{E}\right]
\delta \hat{y}_\alpha\quad \forall \alpha=1,\dots,n\, ,
\end{equation*}
where
\begin{equation*}
    \mathbf{J}_{f}(\vec{s}) = \left(\begin{matrix}
        -\sigma & \sigma & 0\\
        \rho - s_z(t)& -1 & -s_x(t)\\
        s_y(t) & s_x(t) & -\beta
    \end{matrix}\right)\, .
\end{equation*}
In panel (a) of Fig.~\ref{fig:LorenzSynch}, we report the MSF as a function of $\kappa = \epsilon \Lambda^{(\alpha)}$, in the case of the $1-1$ coupling, namely $E_{11}=1$ and all the remaining entries vanish (other examples of linear couplings are briefly presented in Supplementary Note 3). We can observe that the MSF is positive for $\kappa=0$, testifying the chaotic behavior of the reference solution $\vec{s}(t)$, then it steadily decreases and vanishes at $\kappa_1\sim 7.216$. Because the eigenvalues of the Laplace matrix associated to the MWN are given by $\{0,2,4\}$, by taking $2\epsilon > \kappa_1$, we are sure that the MSF is negative (see red circles computed with $\epsilon=4$ in Fig.~\ref{fig:LorenzSynch}(a)) and thus global synchronization is achieved. This can be confirmed by looking at panel (b) of Fig.~\ref{fig:LorenzSynch}, where we show the time evolution of the first component of $\vec{y}_j=\mathbf{R}_\pi \vec{x}_j$ as a function of time. In panel (d) of Fig.~\ref{fig:LorenzSynch}, we report the time evolution of the synchronization error, in the original variables (blue curve): 
\begin{equation*}
    \mathcal{E}(t)=\sqrt{\frac{1}{n(n-1)}\sum_{ij} ||\vec{x}_i(t)-\vec{x}_j(t)||^2}
\end{equation*}
and in the rotated ones (red curve)
\begin{equation*}
    \mathcal{E}_{rot}(t)=\sqrt{\frac{1}{n(n-1)}\sum_{ij} ||\vec{y}_i(t)-\vec{y}_j(t)||^2}\, .
\end{equation*}
One can clearly appreciate that $\mathcal{E}_{rot}(t)\rightarrow 0$ testifying the presence of global synchronization, while the similar quantity, but computed by using the original variables, without rotation, does not decrease in time, wrongly suggesting that the system is out of synchrony.

\begin{figure*}[ht!]
\includegraphics[width=0.8\textwidth]{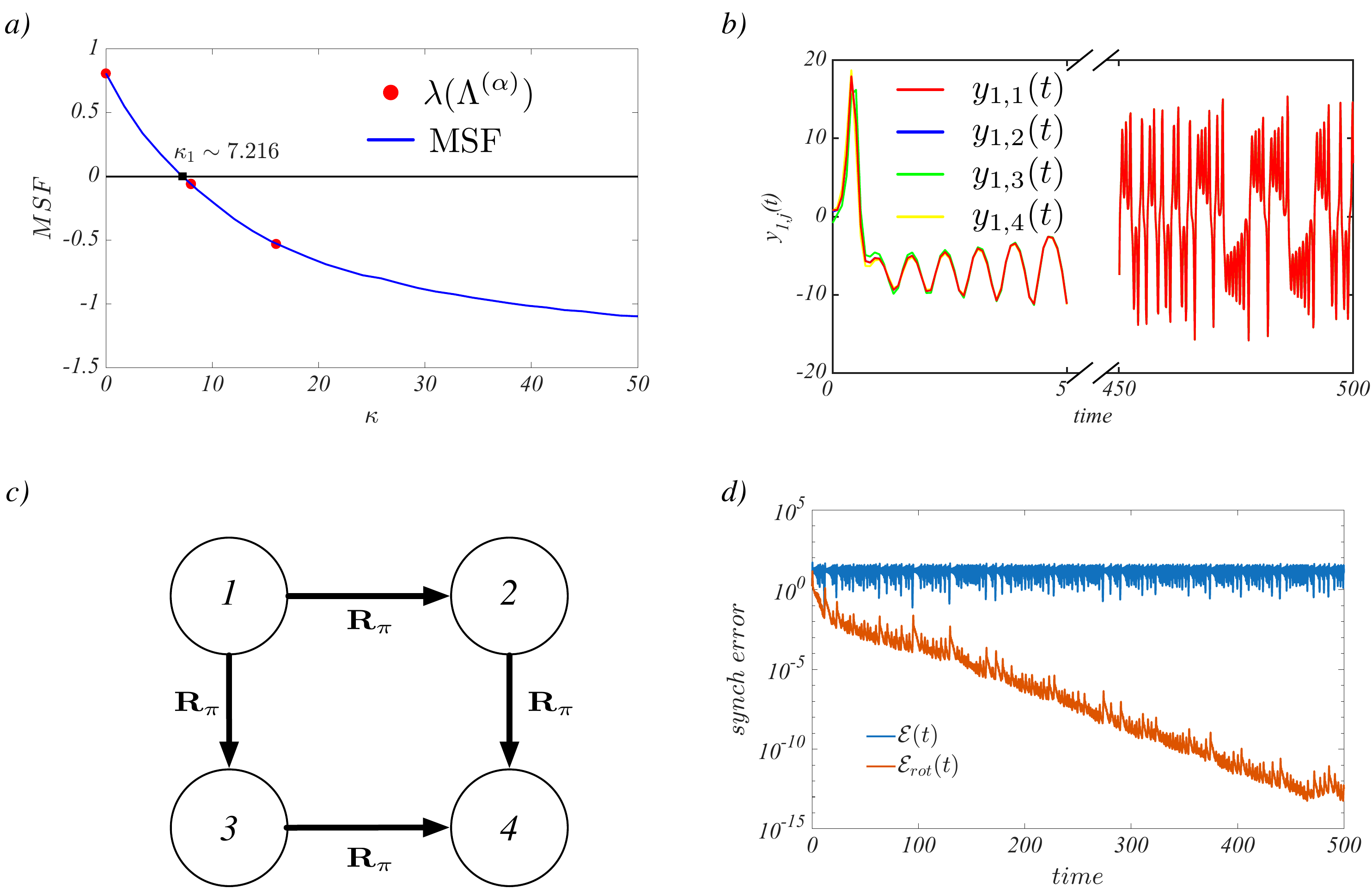}      
\caption{\textbf{Global synchronization of Lorenz oscillators coupled with a square MWN}. 
In panel (a), we report the Master Stability Function, in panel (b) ${y}_{1,j}(t)$, i.e., the first component of $\vec{y}_j$. The used MWN is shown in panel (c), while in panel (d) we report the time evolution of the synchronization error computed in the original $\vec{x}_i$ variables (blue curve) and in the rotated ones, $\vec{y}_i$ (red curve). The coupling parameter is fixed to $\epsilon = 4$.
}
\label{fig:LorenzSynch}
\end{figure*}

\section{Discussion}
\label{sec:dis}
In this work, we have advanced the theoretical understanding of Global Synchronization (GS), a widespread phenomenon underpinning numerous biological rhythms and the reliable operation of engineered systems. Specifically, we have extended existing synchronization theory from systems connected via scalar-weighted links to the more general case of matrix-weighted networks (MWNs). Our focus has been on rotation matrices as edge weights, although the framework could be generalized further to include all orthogonal matrices (notice that to generalize the results for the orthogonal matrices, one needs to impose further constraints on the dynamical systems in response to the invariance of applying these matrices). We have demonstrated that coherence in MWNs is a necessary condition for the emergence of global synchronization. 
Notice that the choice of rotation matrices imposes structural constraints on the class of admissible dynamical systems, namely, only those invariant under the action of these matrices can achieve synchronization. This finding aligns with existing results in the literature and reinforces the intrinsic connection between network topology and dynamical behavior~\cite{Millan2024} to the case of multi-dimensional dynamics.

The existence of GS can be established by using the well-known Master Stability Function (MSF) framework. A central objective of our work has therefore been to extend the MSF formalism to the context of MWNs. Within this framework, we have shown that the spectrum of an appropriately defined Laplacian operator governs the onset of GS. Furthermore, we have identified the existence of a preferred basis, determined by the eigenspace of this Laplacian, in which GS becomes manifest. In other words, while the MSF provides the conditions under which GS occurs, the ability to observe synchronization depends critically on the choice of reference frame. 
This effect has no counterpart in scalar-weighted networks and
represents a novel phenomenon specific to the matrix-weighted setting.
Classical synchronization stability methods, such as the classical MSF, are limited to scalar-weighted networks and assume identical interactions among nodes. These approaches do not easily extend to networks in which node states are multidimensional and edges encode linear transformations between components. In contrast, our framework considers matrix-weighted coupling, explicitly allows interactions involving linear transformations (e.g., rotations), and introduces the notion of coherence. By performing a suitable similarity transformation, the proposed method reduces the system to a generalized Laplacian form while preserving the effect of the transformations. This approach ensures the invariance of the generalized synchronization manifold and enables a rigorous stability analysis in cases where classical scalar-based methods would be insufficient.

The studied framework naturally connects with the broader literature on multilayer and multidimensional networks. In these network structures, {scalar node states across layers or dimensions can be considered as vector states, and the related edges can simultaneously affect such vector states, where inter-layer couplings are commonly assumed to be only between (the copies of) the same nodes.} 
Matrix-weighted couplings provide a principled generalization by allowing explicit linear transformations between the state vectors, enabling collective interactions that {are not emphasized by} 
scalar-weighted multilayer networks. In other words, while traditional multilayer networks restrict how edges influence different components, MWNs unify this perspective by letting each edge fully specify how the state of a source node affects that of a target node. Furthermore, the notion of \textit{coherence} generalizes the idea of layer alignment in multilayer networks, representing a structural feature that is both necessary for synchronization to happen and unique to the MWNs formalism.

Finally, it is worth noting a possible conceptual connection with the H/R (Hopf/Rotation) theorem by Golubitsky and Stewart~\cite{golubitsky1985hopf}, which addresses the emergence of symmetric periodic solutions in networks with equivariant (group-symmetric) dynamics, particularly in the context of pattern formation and bifurcations. While our framework focuses on GS in MWNs rather than bifurcation phenomena, both approaches highlight the key role played by symmetry: in our case, the hypothesis that the nonlinear functions $\vec f$ and $\vec h$ commute with the transformations $\mathbf O_{1i}$ ensures that the generalized synchronization manifold is preserved, recalling the equivariance conditions of the H/R theorem. Exploring this connection in more detail could provide fruitful avenues for future work.

\section{Methods}
\label{sec:meth}

Given the matrix weight of an edge $(i,j)$, $\mathbf{W}_{ij}\in \mathbb{R}^{d\times d}$ {where $d$ is the characteristic dimension of the dynamical system under study}, we rewrite it as $\mathbf{W}_{ij}=w_{ij}\mathbf{R}_{ij}$, where $w_{ij}=||\mathbf{W}_{ij}||_2$ {characterizes the magnitude} and $\mathbf{R}_{ij}\in \mathbb{R}^{d\times d}$ {encodes the transformation with} $||\mathbf{R}_{ij}||_2=1$. In this paper, we focus on the case when each transformation $\mathbf{R}$ is a rotation matrix. We also maintain the same assumption as~\cite{tian2025matrix} where 
$\mathbf{W}_{ij}=\mathbf{W}_{ji}^\top$ for all $i$ and $j$, thus $w_{ij}=w_{ji}$ and $\mathbf{R}_{ij}=\mathbf{R}_{ji}^\top$. Let $d_i=\sum_j w_{ij}$ be the node strength, $\mathcal{D}=\mathbf{D}\otimes \mathbf{I}_d$, the supra-degree matrix, where $\mathbf{D}=\mathrm{diag}(d_1,\dots,d_n)$ and $\mathbf{I}_d$ the $d$-dimensional identity matrix. Then we can eventually define the {\em supra-Laplace matrix}
\begin{equation}
\label{eq:supraL}
\mathcal{L}=\mathcal{D}-\mathcal{W}\, ,
\end{equation}
where the  supra-weight matrix $\mathcal{W}$ has blocks structure, with the $i,j$ block being $\mathbf{W}_{ij}=w_{ij}\mathbf{R}_{ij}$.

The coherence condition is defined by looking at the transformation of each oriented cycle. More precisely, let the oriented cycle be defined by the sequence of distinct edges $(i_1,i_2),(i_2,i_3),\dots,(i_k,i_1)$ and let $\mathbf{R}_{i_j,i_{j+1}}$ be the {transformation} associated to the edge $(i_j,i_{j+1})$, then we require 
\begin{equation}
\label{eq:coherence}
\mathbf{R}_{i_1,i_2}\mathbf{R}_{i_2,i_3}\dots \mathbf{R}_{i_k,i_1}=\mathbf{I}_d\, .
\end{equation}
This allows to define the block diagonal matrix $\mathcal{S}$, whose $i$-th block is the $d\times d$ matrix, $\mathbf{O}_{1i}$, equal to the product of the {transformation of each edge} $\mathbf{O}$ associated to any path starting from nodes in the first block (that can be arbitrarily chosen, hereby set to node $1$) to the block containing node $i$. In formula:
\begin{equation}
\label{eq:matS}
\mathcal{S}=\left(
\begin{matrix}
 \mathbf{I}_d & 0 & \dots & \dots & 0\\
 0 & \mathbf{O}_{12} & \dots & \dots & 0\\
 \vdots & \vdots & \ddots & \vdots & \vdots\\
 0 & \dots & \dots & 0 & \mathbf{O}_{1n}
\end{matrix}\right)\, .
\end{equation}
Let us observe that we introduced above a small abuse of notation by identifying node $i$ with the $i$-th block to which it belongs; this will allow us to simplify the notations and moreover we hereby assume a block to be formed by a single node, differently from the general MWN framework~\cite{tian2025matrix} where a block can contain several nodes, each one connected by edges whose weight is the identity matrix.

We can then prove that given any $\vec{u}\in\mathbf{R}^{d}$, the vector $\vec{v}=\mathcal{S}^\top(\vec{1}_n\otimes \vec{u})$ is an eigenvector of ${\mathcal{L}}$ with eigenvalue $0$ {by Eq.~(2) in~\cite{tian2025matrix}},
that will replace the synchronous manifold in the case of MWN. By using the matrix $\mathcal{S}$ and the supra-Laplacian $\mathcal{L}$, we can define a second supra-Laplacian matrix 
\begin{equation}
\bar{\mathcal{L}} = \mathcal{S}\mathcal{L}\mathcal{S}^\top\, ,
\end{equation}
and we can prove that the latter corresponds to the supra-Laplacian matrix where all the {transformations $\mathbf{O}$} 
have been replaced by the identity matrix $\mathbf{I}_d$ as in~\cite{tian2025matrix}, depending thus only on the scalar weights and the topology of the network:
\begin{equation}
    \label{eq:Lbar}
    \bar{\mathcal{L}}=\bar{\mathbf{L}}\otimes \mathbf{I}_d\, ,
\end{equation}
where $\bar{\mathbf{L}}=\mathbf{D}-\mathbf{A}^{(w)}$, with each $i,j$ element of $\mathbf{A}^{(w)}$ being $w_{ij}$, i.e., $\mathbf{A}^{(w)}$ is the weighted adjacency matrix of the underlying network {in the classic sense}.

We can now state our main result. Namely, let $\vec{s}$ be a reference solution of Eq.~\eqref{eq:dynsys}, then the vector function $\vec{S}(t)$ given by
\begin{equation}
\label{eq:Xsol}
\vec{S}(t) := \mathcal{S}^\top (\vec{1}_n\otimes \vec{s})=\mathcal{S}^\top\left(
\begin{matrix}
 \vec{s}\\ \vec{s}\\ \vdots \\ \vec{s}
\end{matrix}
\right)=\left(
\begin{matrix}
 \vec{s}\\  \mathbf{O}_{12}^\top\vec{s}\\ \vdots \\  \mathbf{O}_{1n}^\top\vec{s}
\end{matrix}
\right)\, ,
\end{equation}
is a solution of Eq.~\eqref{eq:dynsyscoupledLmw2}. To prove this claim, let us compute the left hand side of the Eq.~\eqref{eq:dynsyscoupledLmw2} once we insert $\vec{S}(t)$
\begin{equation}
    \frac{d\vec{S}}{dt}=\mathcal{S}^\top \left(\vec{1}_n\otimes \frac{d\vec{s}}{dt}\right)=\left(
\begin{matrix}
 \frac{d\vec{s}}{dt}\\  \mathbf{O}_{12}^\top\frac{d\vec{s}}{dt}\\ \vdots \\  \mathbf{O}_{1n}^\top\frac{d\vec{s}}{dt}
\end{matrix}
\right)\, ,
\end{equation}
being the matrices $\mathbf{O}_{1i}$ time independent. The right hand side of the same equation, returns
\begin{eqnarray}
 \vec{f}_*(\vec{S})-\mathcal{L}\vec{h}_*(\vec{S})&=&\left(\begin{smallmatrix}
     \vec{f}(\vec{s})\\ \vec{f}(\mathbf{O}^\top_{12}\vec{s})\\\vdots\\ \vec{f}(\mathbf{O}^\top_{1n}\vec{s})
 \end{smallmatrix}\right)-\mathcal{L}\left(\begin{smallmatrix}
     \vec{h}(\vec{s})\\ \vec{h}(\mathbf{O}^\top_{12}\vec{s})\\\vdots\\ \vec{h}(\mathbf{O}^\top_{1n}\vec{s})
\end{smallmatrix}\right)\\
&=&\left(\begin{smallmatrix}
     \vec{f}(\vec{s})\\ \mathbf{O}^\top_{12}\vec{f}(\vec{s})\\\vdots\\ \mathbf{O}^\top_{1n}\vec{f}(\vec{s})
 \end{smallmatrix}\right)-\mathcal{L}\left(\begin{smallmatrix}
     \vec{h}(\vec{s})\\ \mathbf{O}^\top_{12}\vec{h}(\vec{s})\\\vdots\\ \mathbf{O}^\top_{1n}\vec{h}(\vec{s})
 \end{smallmatrix}\right)\notag\\
 &=&\mathcal{S}^\top (\vec{1}_n\otimes \vec{f}(\vec{s}))-\mathcal{L}\mathcal{S}^\top (\vec{1}_n\otimes \vec{h}(\vec{s}))\notag\, ,
\end{eqnarray}
where we used the invariance of $\vec{f}$ and $\vec{h}$ given by~\eqref{eq:condfO}. By observing that $\mathcal{L}\mathcal{S}^\top (\vec{1}_n\otimes \vec{h}(\vec{s}))=0$ and recalling that $\vec{s}$ solves~\eqref{eq:dynsys}, we can thus conclude that
\begin{align}
    \frac{d\vec{S}}{dt}&=\mathcal{S}^\top\left(\vec{1}_n\otimes \frac{d\vec{s}}{dt}\right)\nonumber\\
    &=\mathcal{S}^\top (\vec{1}_n\otimes \vec{f}(\vec{s}))=\vec{f}_*(\vec{S})-\mathcal{L}\vec{h}_*(\vec{S})\, .
\end{align}

To prove the stability of the latter solution $\vec{S}(t)$, we rewrite the $(dn)$-dimensional state vector as $\vec{x}=\vec{S}+\delta\vec{x}$, where $\delta\vec{x}$ is ``small'' perturbation whose time evolution will determine the stability of $\vec{S}$: if $\delta\vec{x}$ converges to zero, then $\vec{x}$ will converge to $\vec{S}$ and thus the system will exhibit GS. Otherwise, if $\delta\vec{x}$ will stay away from zero, then GS cannot be achieved because $\vec{x}$ will deviate from $\vec{S}$.

To study the time evolution of $\delta\vec{x}$ we expand to the first order system~\eqref{eq:dynsyscoupledLmw2} to get 
\begin{equation}
\label{eq:lin1}
\frac{d\delta\vec{x}}{dt}=\mathbf{J}_{f_*}(\vec{S})\delta\vec{x}-\mathcal{L}\mathbf{J}_{h_*}(\vec{S})\delta\vec{x}\, ,
\end{equation}
where $\mathbf{J}_{f_*}(\vec{S})$ and $\mathbf{J}_{h_*}(\vec{S})$ are respectively the Jacobian of ${f_*}$ and ${h_*}$ evaluated on the solution $\vec{S}$. The latter can be written in ``components'' by defining $\delta\vec{x}=(\delta x_1^\top,\dots,\delta x_n^\top)^\top$ and to obtain Eq.~\eqref{eq:lin2}, hereby rewritten to help the reader
\begin{equation}
\frac{d}{dt}
 \delta \vec{x}_j=\mathbf{J}_{f}(\mathbf{O}_{1j}^\top\vec{s})
 \delta \vec{x}_j-\sum_\ell \mathcal{L}_{j\ell}\mathbf{J}_{h}(\mathbf{O}_{1\ell}^\top\vec{s})
 \delta \vec{x}_\ell\, ,
\end{equation}

The invariance condition~\eqref{eq:condfO} translates into a similar one for the Jacobian matrices, more precisely, it allows to obtain
\begin{equation}
 \mathbf{J}_{f}(\mathbf{O}_{1j}^\top\vec{s})=\mathbf{O}_{1j}^\top\mathbf{J}_{f}(\vec{s})\mathbf{O}_{1j}
\end{equation}
and
\begin{equation}\mathbf{J}_{h}(\mathbf{O}_{1j}^\top\vec{s})=\mathbf{O}_{1j}^\top\mathbf{J}_{h}(\vec{s})\mathbf{O}_{1j}\, .
\end{equation}
Hence, we can conclude that
\begin{equation}
\frac{d}{dt}
 \delta \vec{x}_j=\mathbf{O}_{1j}^\top\mathbf{J}_{f}(\vec{s})\mathbf{O}_{1j}
 \delta \vec{x}_j-\sum_\ell \mathcal{L}_{j\ell}\mathbf{O}_{1\ell}^\top\mathbf{J}_{h}(\vec{s})
 \mathbf{O}_{1\ell}\delta \vec{x}_\ell\, ,
\end{equation}
or equivalently by defining $\vec{y}_j=\mathbf{O}_{1j}\vec{x}_j$
\begin{eqnarray}
\frac{d}{dt}
 \delta \vec{y}_j&=&\mathbf{J}_{f}(\vec{s})
 \delta \vec{y}_j-\sum_\ell \mathbf{O}_{1j}\mathcal{L}_{j\ell}\mathbf{O}_{1\ell}^\top\mathbf{J}_{h}(\vec{s})
 \delta \vec{y}_\ell\notag\\
 &=&\mathbf{J}_{f}(\vec{s})
 \delta \vec{y}_j-\sum_\ell \mathcal{{L}}_{j\ell}\mathbf{J}_{h}(\vec{s})
 \delta \vec{y}_\ell\, ,
\end{eqnarray}
namely Eq.~\eqref{eq:lin3} in the main text. Let us observe that to get the latter formula we made use of the definition of the supra-Laplace matrix $\bar{\mathcal{L}}$ whose entries are related to the Laplace matrix of the weighted underlying network~\eqref{eq:Lbar}. Let us observe that by using the stack vectors $\vec{x}$ and $\vec{y}$, the above change of variables can be rewritten as $\vec{y}=\mathcal{S}\vec{x}$.

To prove the stability of $\delta \vec{y}_j$ and hence of $\delta \vec{x}_j$, we exploit the existence of an orthonormal basis for the Laplace matrix $\bar{\mathbf{L}}$, i.e., $\bar{\phi}^{(\alpha)}$, $\Lambda^{(\alpha)}$, $\alpha=1,\dots,n$, to project $\delta \vec{y}_j$ onto the latter
\begin{equation}
\delta \vec{y}_j=\sum_\alpha
\delta \hat{y}_\alpha\bar{\phi}^{(\alpha)}_j\, .
\end{equation}
In this way, Eq.~\eqref{eq:lin3} returns
\begin{align}
\label{eq:lin4}
\sum_\alpha\frac{d}{dt}
\delta \hat{y}_\alpha\bar{\phi}^{(\alpha)}_j=\mathbf{J}_{f}(\vec{s})&\sum_\alpha
\delta \hat{y}_\alpha\bar{\phi}^{(\alpha)}_j\nonumber\\
&- \sum_\alpha \Lambda^{(\alpha)}\mathbf{J}_{h}(\vec{s})
\delta \hat{y}_\alpha\bar{\phi}^{(\alpha)}_j\, .
\end{align}
By left multiplying by $\bar{\phi}^{(\alpha)}$ and by using the orthonormality of eigenvectors we eventually obtain Eq.~\eqref{eq:lin5}, hereby reported:
\begin{eqnarray}
\frac{d}{dt}
\delta \hat{y}_\alpha &=& \left[\mathbf{J}_{f}(\vec{s})-\Lambda^{(\alpha)}\mathbf{J}_{h}(\vec{s})\right]
\delta \hat{y}_\alpha \quad \forall \alpha=1,\dots,n\notag\\
&=:&\mathbf{J}_{\alpha}\delta \hat{y}_\alpha\, .
\end{eqnarray}

This is a non-autonomous linear system containing the information about the dynamics and the coupling via the Jacobian matrices, while the MWN enters only via the eigenvalues of the supra-Laplace matrix $\bar{\mathcal{L}}$ depending only on the scalar weights. Recall however that the rotation weights play a central role in determining the coherence condition. Global synchronization is thus obtained by proving that for all $\Lambda^{(\alpha)}$ the largest Lyapunov exponent of Eq.~\eqref{eq:lin5} is negative. In the case the latter system is autonomous, this accounts to prove that the real part of the spectrum of $\mathbf{J}_{\alpha}$ is negative for all $\alpha$.

\section*{Acknowledgements}
R.L. acknowledges support from the EPSRC Grants EP/V013068/1, EP/V03474X/1, and EP/Y028872/1.

\section*{Author contributions}
A.G., Y.T., R.L., T.C. contributed equally to the project and the preparation of the manuscript.

\section*{Data Availability}
Data sharing not applicable to this article as no datasets were generated or analysed during the current study.

\section*{Code Availability}
The codes supporting the findings of this study are available from the authors upon request.

\section*{Competing interests}
The authors declare no competing interests.

\bibliographystyle{unsrt}
\bibliographystyle{abbrv}
\bibliography{mybib}

\clearpage

\onecolumngrid

\appendix


\renewcommand{\figurename}{\textbf{Supplementary Figure}}
\renewcommand{\tablename}{\textbf{Supplementary Table}}
\makeatletter
\renewcommand\@biblabel[1]{#1.}
\renewcommand{\bibsection}{\section*{\refname}}
\makeatother
\renewcommand{\refname}{Supplementary References}

\section*{Supplementary Note 1:\\ Analysis of the Stuart-Landau model defined on MWN}
\label{sec:appSL}

The Stuart-Landau (SL) model~\cite{Stuart1978,vanharten,aranson,garcamorales} can be written in Cartesian coordinates as follows 
\begin{equation}
\label{eq:appSLmat}
\frac{d}{dt}
\begin{pmatrix}
 x\\y
\end{pmatrix} =
\begin{pmatrix}
 \sigma_{\Re} & -\sigma_{\Im}\\
 \sigma_{\Im} & \sigma_{\Re}
\end{pmatrix}
\begin{pmatrix}
 x\\y
\end{pmatrix}\\ -  (x^2+y^2)
\begin{pmatrix}
 \beta_{\Re} & -\beta_{\Im}\\
 \beta_{\Im} & \beta_{\Re}
\end{pmatrix}
\begin{pmatrix}
 x\\y
\end{pmatrix}\,,
\end{equation}
with complex model parameters $\sigma=\sigma_{\Re}+i\sigma_{\Im}$ and $\beta=\beta_{\Re}+i\beta_{\Im}$. By exploiting the invariance by rotation, one can look for a solution in polar coordinates, $x(t)=r(t)\cos(\theta(t))$ and $y(t)=r(t)\sin(\theta(t))$, and thus obtain for $\sigma_{\Re}>0$ and $\beta_{\Re}>0$ the stable limit cycle solution $x_{LC}(t)=\sqrt{\sigma_{\Re}/\beta_{\Re}}\cos(\omega t)$, $y_{LC}(t)=\sqrt{\sigma_{\Re}/\beta_{\Re}}\sin(\omega t)$, where $\omega =\sigma_{\Im}-\beta_{\Im} \sigma_{\Re}/\beta_{\Re}$. 

Let us, now, assume to have $n$ identical SL oscillators~\eqref{eq:appSLmat} anchored to the nodes of a Weighted Matrix Networks (MWN) and coupled via a diffusive-like nonlinear function; hence, the $j$-th oscillator evolves according to 
\begin{align}
\label{eq:appSLmatnet}
\frac{d}{dt}
\begin{pmatrix}
 x_j\\y_j
\end{pmatrix} &=
\begin{pmatrix}
 \sigma_{\Re} & -\sigma_{\Im}\\
 \sigma_{\Im} & \sigma_{\Re}
\end{pmatrix}
\begin{pmatrix}
 x_j\\y_j
\end{pmatrix} -  (x_j^2+y_j^2)
\begin{pmatrix}
 \beta_{\Re} & -\beta_{\Im}\\
 \beta_{\Im} & \beta_{\Re}
\end{pmatrix}
\begin{pmatrix}
 x_j\\y_j
\end{pmatrix}-\sum_\ell \mathcal{L}_{j\ell}\left[ (x_\ell^2+y_\ell^2)^{\frac{m-1}{2}}
\begin{pmatrix}
 \mu_{\Re} & -\mu_{\Im}\\
 \mu_{\Im} & \mu_{\Re}
\end{pmatrix}
\begin{pmatrix}
 x_\ell\\y_\ell
\end{pmatrix}\right]\nonumber\\
&=:\vec{f}(x_j,y_j)-\sum_\ell \mathcal{L}_{j\ell}\vec{h}(x_\ell,y_\ell)\quad \forall j=1\dots,n\, ,
\end{align}
where $\vec{h}(x_\ell,y_\ell):=(x_\ell^2+y_\ell^2)^{\frac{m-1}{2}}
\begin{pmatrix}
 \mu_{\Re} & -\mu_{\Im}\\
 \mu_{\Im} & \mu_{\Re}
\end{pmatrix}
\begin{pmatrix}
 x_\ell\\y_\ell
\end{pmatrix}$, $\mu=\mu_{\Re}+i\mu_{\Im}$ is the complex coupling strength and  $\mathcal{L}=\mathcal{D}-\mathcal{W}$ is the supra-Laplace matrix associated with the MWN~\cite{tian2025matrix}, being $\mathcal{W}$ the supra-adjacency matrix whose $(ij)$-th block is the matrix $\mathbf{W}_{ij}=w_{ij}\mathbf{R}_{ij}$, with scalar weight $w_{ij}>0$ and $2\times 2$ rotation matrix $\mathbf{R}_{ij}$. And $\mathcal{D}=\mathbf{D}\otimes \mathbf{I}_2$ the $(2n) \times (2n)$ supra-degree matrix obtained via the $2$-dimensional identity matrix, $\mathbf{I}_2$, and the degree matrix $\mathbf{D}=\mathrm{diag}(d_1,\dots,d_n)$, with degree (strength) $d_i=\sum_j w_{ij}$.

If the MWN is coherent, then $\vec{S}(t)= \mathcal{S}^\top (\vec{1}_n\otimes \vec{s}_{LC}(t))$ is a solution of~\eqref{eq:appSLmatnet}. Such a solution is obtained by combining the limit cycle $\vec{s}_{LC}(t)=(x_{LC}(t),y_{LC}(t))$ describing the evolution of every single node and the block-diagonal matrix
\begin{equation}
\label{eq:appmatS}
\mathcal{S}=\left(
\begin{matrix}
 \mathbf{I} & 0 & \dots & \dots & 0\\
 0 & \mathbf{O}_{12} & \dots & \dots & 0\\
 \vdots & \vdots & \ddots & \vdots & \vdots\\
 0 & \dots & \dots & 0 & \mathbf{O}_{1n}
\end{matrix}\right)\, ,
\end{equation}
where $\mathbf{O}_{1\ell}$ is the transformation of any oriented path going from node $1$ to any node in the $\ell$-th block. Notice that the construction of such a matrix stems from verifying the coherence property~\cite{tian2025matrix}.

The proof of the latter claim follows the same reasoning as the general argument presented in the main text and will therefore not be repeated here. However, we will now develop in detail the computation required to prove the stability of the solution, which ensures the existence of Global Synchronization (GS) of SL oscillators coupled with a MWN.

To determine the conditions allowing for the stability of $\vec{S}=\mathcal{S}^\top (\vec{1}_n\otimes \vec{s}_{LC})$, we perform a linear stability analysis of~\eqref{eq:appSLmatnet} close to the latter solution; we thus write the $j$-th ``component'' of the $2n-$dimensional state vector $\vec{x}$ as $\vec{x}_j=\mathbf{O}_{1j}^\top\vec{s}+\delta\vec{x}_j$, where $\delta\vec{x}_j=(\delta x_j,\delta y_j)^\top$ is a ``small'' perturbation whose time evolution will determine the stability of $\vec{S}$. By plugging the above ansatz into Eq.~\eqref{eq:appSLmatnet} and using the fact that $\vec{s}$ is a solution, we end up with the linearized dynamics at first order in $\delta \vec{x}_j$:

\begin{equation}
\label{eq:applin2}
\frac{d}{dt}\left(
\begin{matrix}
 \delta x_j\\ \delta y_j
\end{matrix}
\right)=\mathbf{J}_{f}(\mathbf{O}_{1j}^\top\vec{s})\left(
\begin{matrix}
 \delta x_j\\ \delta y_j
\end{matrix}
\right)-\sum_\ell \mathcal{L}_{j\ell}\mathbf{J}_{h}(\mathbf{O}_{1\ell}^\top\vec{s})\left(
\begin{matrix}
 \delta x_\ell\\ \delta y_\ell
\end{matrix}
\right)\, .
\end{equation}
Moreover, invariance of the SL system and of the coupling function with respect to the matrices $\mathbf{O}_{1,\ell}$ allows us to obtain
\begin{equation}
\frac{d}{dt}\left(
\begin{matrix}
 \delta x_j\\ \delta y_j
\end{matrix}
\right)=\mathbf{O}_{1j}^\top\mathbf{J}_{f}(\vec{s})\mathbf{O}_{1j}\left(
\begin{matrix}
 \delta x_j\\ \delta y_j
\end{matrix}
\right)-\sum_\ell \mathcal{L}_{j\ell}\mathbf{O}_{1\ell}^\top\mathbf{J}_{h}(\vec{s})\mathbf{O}_{1\ell}\left(
\begin{matrix}
 \delta x_\ell\\ \delta y_\ell
\end{matrix}
\right)\, .
\end{equation}
Eventually, by introducing $\left(
\begin{matrix}
 \delta u_j\\ \delta v_j
\end{matrix}
\right)=\mathbf{O}_{1j}\left(
\begin{matrix}
 \delta x_j\\ \delta y_j
\end{matrix}
\right)$ we get
\begin{equation}
\label{eq:applin3}
\frac{d}{dt}\left(
\begin{matrix}
 \delta u_j\\ \delta v_j
\end{matrix}
\right)=\mathbf{J}_{f}(\vec{s})\left(
\begin{matrix}
 \delta u_j\\ \delta v_j
\end{matrix}
\right)-\sum_\ell \bar{{L}}_{j\ell}\mathbf{J}_{h}(\vec{s})\left(
\begin{matrix}
 \delta u_\ell\\ \delta v_\ell
\end{matrix}
\right)\, ,
\end{equation}
where we introduced the Laplace matrix $\bar{\mathbf{L}}$ obtained from the supra-Laplace matrix $\bar{\mathcal{L}} = \mathcal{S}\mathcal{L}\mathcal{S}^\top$, and depending only on the scalar weights. The largest Lyapunov exponent of the previous system allows us to inquire about the stability of the solution $\vec{S}$; however, this system has the same size as the MWN and thus it can be very large. To overcome this issue, we follow the approach proposed first by Fujisaka~\cite{fujisaka1983stability} and Pecora and collaborators~\cite{Pecora_etal97,pecora1998master}, which allows one to deal with a $1$-parameter family of much smaller systems, thus being able to improve our understanding of the model.

Let $\bar{\phi}^{(\alpha)}$, $\Lambda^{(\alpha)}$, $\alpha=1,\dots,n$, be orthogonal eigenvectors and nonnegative eigenvalues of $\bar{\mathbf{L}}$, then we can then project $(\delta u_j,\delta v_j)^\top$ onto the eigenbasis and obtain

\begin{equation}
\left(
\begin{matrix}
\delta u_j\\ \delta v_j 
\end{matrix}
\right)=\sum_\alpha\left(
\begin{matrix}
\delta \hat{u}_\alpha\\ \hat{\delta v}_\alpha 
\end{matrix}
\right)\bar{\phi}^{(\alpha)}_j\, ,
\end{equation}
that plugged in Eq.~\eqref{eq:applin3} returns
\begin{equation}
\label{eq:applin4}
\sum_\alpha\frac{d}{dt}\left(
\begin{matrix}
\delta \hat{u}_\alpha\\ \hat{\delta v}_\alpha 
\end{matrix}
\right)\bar{\phi}^{(\alpha)}_j=\mathbf{J}_{f}(\vec{s})\sum_\alpha\left(
\begin{matrix}
\delta \hat{u}_\alpha\\ \hat{\delta v}_\alpha 
\end{matrix}
\right)\bar{\phi}^{(\alpha)}_j- \sum_\alpha \Lambda^{(\alpha)}\mathbf{J}_{h}(\vec{s})\left(
\begin{matrix}
\delta \hat{u}_\alpha\\ \hat{\delta v}_\alpha 
\end{matrix}
\right)\bar{\phi}^{(\alpha)}_j\, .
\end{equation}
Using the orthonormality assumption on the eigenvectors allows to conclude

\begin{equation}
\label{eq:applin5}
\frac{d}{dt}\left(
\begin{matrix}
\delta \hat{u}_\alpha\\ \hat{\delta v}_\alpha 
\end{matrix}
\right) = \left[\mathbf{J}_{f}(\vec{s})-\Lambda^{(\alpha)}\mathbf{J}_{h}(\vec{s})\right]\left(
\begin{matrix}
\delta \hat{u}_\alpha\\ \hat{\delta v}_\alpha 
\end{matrix}
\right) \quad \forall \alpha=1,\dots,n\, ,
\end{equation}
that is the Master Stability Equation for the Weighted Matrix Network framework.

In the case of the SL model and the nonlinear coupling given by $\vec{h}(x_\ell,y_\ell):=(x_\ell^2+y_\ell^2)^{\frac{m-1}{2}}\left(
\begin{smallmatrix}
 \mu_{\Re} & -\mu_{\Im}\\
 \mu_{\Im} & \mu_{\Re}
\end{smallmatrix}\right)\left(
\begin{smallmatrix}
 x_\ell\\y_\ell
\end{smallmatrix}\right)$, one can explicitly compute the Jacobian matrices which turn out to be constant provided we use the ansatz $u_j=x_{LC}+x_{LC}\delta u_j-y_{LC}\delta v_j$ and $v_j=y_{LC}+x_{LC}\delta v_j+y_{LC}\delta u_j$. More precisely, we get
\begin{equation}
\label{eq:apprhotheta}
\frac{d}{dt}\left(
\begin{matrix}
 \delta\hat{u}_\alpha\\ \delta\hat{v}_\alpha
\end{matrix}\right)=\left[ \left(
\begin{matrix}
 -2\sigma_{\Re} & 0\\
 -2\beta_{\Im} \frac{\sigma_{\Re}}{\beta_{\Re}} &0
\end{matrix}
\right)-\left(\frac{\sigma_{\Re}}{\beta_{\Re}}\right)^{\frac{m-1}{2}}\Lambda^{(\alpha)}\left(
\begin{matrix}
 m\mu_{\Re} & -\mu_{\Im}\\
 m\mu_{\Im} & \mu_{\Re}
\end{matrix}
\right)\right]\left(
\begin{matrix}
\delta\hat{u}_\alpha\\ \delta\hat{v}_\alpha
\end{matrix}\right)=:\mathbf{J}_\alpha \left(
\begin{matrix}
\delta\hat{u}_\alpha\\ \delta\hat{v}_\alpha
\end{matrix}\right)\, .
\end{equation}

Let, now, $\lambda(\Lambda^{(\alpha)}) = \max_{j} \Re\lambda_j (\Lambda^{(\alpha)})$ be the characteristic root of $\mathbf{J}_\alpha=\mathbf{J}_{f}(\vec{s})-\Lambda^{(\alpha)}\mathbf{J}_{h}(\vec{s})$ with the largest real part, this is the celebrated Master Stability Function, allowing to infer the stability or instability of the synchronized solution $\vec{S}$ as a function of the spectrum $\Lambda^{(\alpha)}$ of the underlying network. More precisely, if for all $\alpha$ we get $\lambda(\Lambda^{(\alpha)})<0$, then $\vec{S}$ is stable and the system synchronizes, i.e., the solution $\vec{x}$, starting close enough to $\vec{S}$, will converge to the latter as time increases. On the other hand, if there exists $\alpha >1$ for which $\lambda(\Lambda^{(\alpha)})>0$, then $\vec{S}$ is unstable and the system does not synchronize.

\section*{Supplementary Note 2:\\ Analysis of the $3$D-rotation model defined on MWN}
\label{sec:appeqstab3D}

The aim of this section is to provide more details about the ``isolated'' system in $3-$dimension and its counterpart defined on a MWN. Let us recall that the former system is

\begin{equation}\label{eq:3Dmodisolapp}
\frac{d{\vec{x}}}{dt}=\mathbf{A}\vec{x}-|\vec{x}|^2\mathbf{B}\vec{x}\, ,
\end{equation}
where $\vec{x}\in\mathbb{R}^3$ and $\mathbf{A}$ and $\mathbf{B}$ are $3\times 3$ matrices 
\begin{align*}
\mathbf{A}=
\begin{pmatrix}
\mathbf{A}_1 & \vec{0}\\
\vec{0}^\top & \mu_A 
\end{pmatrix}\text{ and }\mathbf{B}=
\begin{pmatrix}
\mathbf{B}_1 & \vec{0}\\
\vec{0}^\top &\mu_B 
\end{pmatrix},
\end{align*}
with $\mu_A, \mu_B\in\mathbb R$, $\vec{0}=(0,0)^\top$. $\mathbf A_1$ and $\mathbf B_1$ are given by
\begin{align*}
\mathbf{A}_1=\lambda_A
\begin{pmatrix}
\cos(a) & -\sin(a)\\
\sin(a) & \cos(a)
\end{pmatrix}\text{ and }
\mathbf{B_1}=\lambda_B
\begin{pmatrix}
\cos(b) & -\sin(b)\\
\sin(b) & \cos(b)
\end{pmatrix}.
\end{align*}

The special form of the above matrices suggests decomposing any vector of $\mathbb{R}^3$ as
\begin{equation*}
\vec{x}=\vec{x}^{||}+\vec x^{\perp}\equiv \left(
\begin{matrix}
\vec{\xi} \\0
\end{matrix}
\right)+\left(
\begin{matrix}
\vec{0}\\x_3
\end{matrix}
\right)\, , \quad\text{ with } \vec{\xi}=(\xi_1,\xi_2)^\top\in\mathbb{R}^2 \text{ and } x_3\in\mathbb{R}\, ,
\end{equation*}
where $\vec{x}^{||}$ is the component lying in the rotation plane left invariant by $\mathbf{A}_1$ and $\mathbf{B}_1$, while $\vec{x}^{\perp}$ lies in the direction orthogonal to such plane. Thus, system~\eqref{eq:3Dmodisolapp} can be rewritten as
\begin{equation}
\begin{cases}
 \frac{d{\vec{\xi}}}{dt} &=\mathbf{A}_1\vec{\xi}-\left(|\vec{\xi}|^2+x_3^2\right) \mathbf{B}_1\vec{\xi},\\
 \frac{d{x}_3}{dt}&=\mu_A x_3-\left(|\vec{\xi}|^2+x_3^2\right)\mu_B x_3\, .
\end{cases}
\end{equation}

To make a step forward, let us introduce polar coordinates for $\vec{\xi}$, namely
$\xi_1=r\cos(\phi)$ and $\xi_2=r\sin(\phi)$,
to eventually obtain
\begin{equation}
\label{eq:3Dmodisolter}
\begin{cases}
 \frac{d{r}}{dt} &=\left[\lambda_A \cos(a) -(r^2+x_3^2)\lambda_B\cos(b)\right]r,\\
 \frac{d{\phi}}{dt}&=\lambda_A \sin(a) -(r^2+x_3^2)\lambda_B\sin(b),\\
 \frac{d{x}_3}{dt}&=\left[\mu_A -\left(r^2+x_3^2\right)\mu_B\right] x_3\, ,
\end{cases}
\end{equation}

Let us now determine the equilibria of the latter system and their stability. By direct inspection, we realize the existence of four equilibria for $r$ and $x_3$, while $d{\phi}/dt$ will not (generically) vanish and thus the angle $\phi$ will linearly evolve in time:
\begin{enumerate}
\item $x_3=r=0$;
\item $x_3=0$ and $r=\sqrt{\frac{\lambda_A \cos(a)}{\lambda_B \cos(b)}}$, provided $\frac{\lambda_A \cos(a)}{\lambda_B \cos(b)}>0$;
\item $x_3=\sqrt{\frac{\mu_A}{\mu_B}}$ and $r=0$, provided $\frac{\mu_A}{\mu_B}>0$;
\item $x_3^2+r^2=\frac{\mu_A}{\mu_B}=\frac{\lambda_A\cos(a)}{\lambda_B\cos(b)}$. Notice that in this case we have a $1$-parameter family of equilibria, being $x_3$ and $r$ not fixed.
\end{enumerate}

Notice that the Jacobian matrix of~\eqref{eq:3Dmodisolter} is given by
\begin{align}
\label{eq:jacobian}
\mathbf{J}=
\begin{pmatrix}
 \lambda_A\cos(a)-(3r^2+x_3^2)\lambda_B\cos(b) & 0 & -2x_3r\lambda_B\cos(b)\\
 -2r\lambda_B\sin(b) & 0 & -2x_3 \lambda_B\sin(b)\\
 -2r\mu_Bx_3 & 0 & \mu_A-(r^2+3x_3^2)\mu_B
\end{pmatrix}\, .
\end{align}
Because of the special structure of such a matrix, one can easily compute its eigenvalues and thus conclude that:
\begin{enumerate}
\item the equilibrium $x_3=r=0$ is stable if 
\begin{align*}
    \lambda_A\cos(a)<0 \text{ and } \mu_A<0;
\end{align*}
\item the equilibrium 
$x_3=0$, $r=\sqrt{\frac{\lambda_A \cos(a)}{\lambda_B \cos(b)}}$, exists and it is stable provided 
\begin{equation}
    \label{eq:appcondstab}
    \lambda_A\cos(a)>0\, , \lambda_B\cos(b)>0 \text{ and }
\mu_A\lambda_B\cos(b)-\mu_B\lambda_A\cos(a)<0\, ;
\end{equation}
\item the equilibrium $x_3=\sqrt{\frac{\mu_A}{\mu_B}}$, $r=0$, exists and it is stable provided $\mu_A>0$, $\mu_B>0$, and
$\mu_A\lambda_B\cos(b)-\mu_B\lambda_A\cos(a)>0$;
\item The equilibrium $x_3^2+r^2=\frac{\mu_A}{\mu_B}=\frac{\lambda_A\cos(a)}{\lambda_B\cos(b)}$ exists and it is stable provide $\mu_A\lambda_B\cos(b)-\mu_B\lambda_A\cos(a)=0$ and $x_3^2\mu_B+r^2\lambda_B\cos(b)>0$.
\end{enumerate}

Let us now consider $n$ identical copies of system~\eqref{eq:3Dmodisolapp} coupled via a Matrix Weighted Network; we are thus dealing with
\begin{align}
\frac{d{\vec{x}}_i}{dt}=\vec{f}(\vec{x}_i)-\epsilon \mathbf{C} \sum_j \mathcal{L}_{ij}\vec{x}_j
=\mathbf{A}\vec{x}_i-|\vec{x}_i|^2\mathbf{B}\vec{x}_i-\epsilon \mathbf{C} \sum_j \mathcal{L}_{ij}\vec{x}_j,\quad\forall j\in\{1,\dots,n\},
\end{align}
where $\vec{x}_i\in\mathbb{R}^3$, $\mathbf{A}$ and $\mathbf{B}$ are $3\times 3$ matrices with the same properties of those defined above and $\mathbf{C}$ is a third $3\times 3$ matrix with a similar structure
\begin{align*}
\mathbf{C}=
\begin{pmatrix}
\mathbf{C}_1 & \vec{0}\\
\vec{0}^\top & \mu_C 
\end{pmatrix}
\equiv 
\begin{pmatrix}
\lambda_C 
\begin{pmatrix}
\cos(c) & -\sin(c)\\
\sin(c) & \cos(c)
\end{pmatrix}
 & \vec{0}\\
\vec{0}^\top & \mu_C 
\end{pmatrix},
\end{align*}
with $\lambda_C$, $\mu_C\in\mathbb R$.

By assuming the MWN to be coherent and with matrix weights to be rotation about the direction $\vec{u}=(0,0,1)^\top$, then we can perform the change of coordinates $\vec{y}_i=\mathbf{O}_{1i}\vec{x}_i$ for all $i=1,\dots,n$ to get 
\begin{align}
\label{eq:3Dmodnetyapp}
\frac{d{\vec{y}}_i}{dt}=\mathbf{A}\vec{y}_i-|\vec{y}_i|^2\mathbf{B}\vec{y}_i-\epsilon \mathbf{C} \sum_j \bar{{L}}_{ij}\vec{y}_j \quad\forall j\in\{1,\dots,n\}\, .
\end{align}
By exploiting, once again, the splitting induced by the form of the rotation matrices, we can define 
\begin{align*}
\vec{y}_i=\vec{y}_i^{||}+\vec y_i^{\perp}\equiv
\begin{pmatrix}
\vec{\xi}_i \\0
\end{pmatrix}
+
\begin{pmatrix}
\vec{0}\\y_{3,i}
\end{pmatrix},
\end{align*}
and obtain
\begin{align}
\label{eq:applinsyst}
\begin{cases}
 \frac{d{\vec{\xi}}_j}{dt} =\mathbf{A}_1\vec{\xi}_j-\left(|\vec{\xi}_j|^2+y_{3,j}^2\right) \mathbf{B}_1\vec{\xi}_j-\epsilon \mathbf{C} \sum_\ell \bar{{L}}_{j\ell}\vec{\xi}_\ell,\\
 \frac{d{y}_{3,j}}{dt}=\mu_A y_{3,j}-\left(|\vec{\xi}_j|^2+y_{3,j}^2\right)\mu_B y_{3,j}-\epsilon \mu_{C} \sum_\ell \bar{{L}}_{j\ell}\vec{y}_{3,\ell}\, .
\end{cases}
\end{align}

Under the conditions, $\lambda_A\cos(a)>0$, $\lambda_B\cos(b)>0$ and $
\mu_A\lambda_B\cos(b)-\mu_B\lambda_A\cos(a)<0$, the  ``isolated'' system will admit the stable equilibrium $\hat{x}_3=0$, $\hat{r}=\sqrt{\frac{\lambda_A \cos(a)}{\lambda_B \cos(b)}}$ and $d{\phi}/dt = \lambda_A\frac{\sin(a-b)}{\cos(b)}=\omega$. We are interested to prove that the system composed by $n$ identical copies coupled via a MWN is capable to globally synchronize to $\vec{S}(t)=\mathcal{S}^\top(\vec{1}_n\otimes \vec{s}(t))$ where $\vec{s}(t)= \left(\hat{r}\cos(\omega  t),\hat{r}\sin(\omega t),0\right)^\top\equiv (\hat{\xi}_1(t),\hat{\xi}_2(t),0)^\top$.

Because the MWN satisfies the coherence assumption and, moreover, the dynamical system preserves the latter, to prove the onset of GS, it remains to show that $\vec{S}(t)$ is stable. Let us write $\vec{\xi}_j=(\xi_{1,j},\xi_{2,j})^\top$, and define a perturbation of the reference solution in the following form

\begin{eqnarray*}
\begin{cases}
\xi_{1,j}&=\hat{\xi}_1+\hat{\xi}_1\delta u_j-\hat{\xi}_2\delta v_j\\
\xi_{2,j}&=\hat{\xi}_2+\hat{\xi}_1\delta v_j+\hat{\xi}_2\delta u_j\\
y_{3,j}&=\delta \eta_j\, ,
\end{cases}
\end{eqnarray*}
where $\delta u_j$, $\delta v_j$ and $\delta \eta_j$ are ``small'' functions of time. Inserting the latter definition into~\eqref{eq:applinsyst} and neglecting all the terms of order larger than one, we obtain

\begin{align}
\label{eq:app3Dmodnetyquat}
\begin{dcases}
  \frac{d\delta{u}_j}{dt} = -2\lambda_A \cos(a) \delta u_j -\epsilon \lambda_C \sum_\ell \bar{L}_{j\ell} \left( \cos(c) \delta u_\ell-\sin(c) \delta v_\ell\right)\\
 \frac{d\delta{v}_j}{dt} = -2\lambda_A \frac{\cos(a)\sin(b)}{\cos(b)}\delta u_j -\epsilon \lambda_C \sum_\ell \bar{L}_{j\ell} \left( \sin(c) \delta u_\ell+\cos(c) \delta v_\ell\right)\\
 \frac{d\delta{\eta}_j}{dt}=\left(\mu_A -\frac{\lambda_A \cos(a)}{\lambda_B \cos(b)}\mu_B\right)\delta\eta_j-\epsilon \mu_C \sum_\ell \bar{L}_{j\ell} \delta\eta_{\ell}\, .
\end{dcases}
\end{align}
Let $\bar{\phi}^{(\alpha)}$, $\Lambda^{(\alpha)}$, $\alpha=1,\dots,k$, the orthogonal eigenvectors and nonnegative eigenvalues of $\bar{\mathbf{L}}$. Then we can project $\delta u_j$, $\delta v_j$ and $\delta\eta_j$ onto this basis

\begin{align}
\delta u_j=\sum_\alpha \delta u_\alpha  \bar{\phi}_j^{(\alpha)},\quad \delta v_j=\sum_\alpha \delta v_\alpha  \bar{\phi}_j^{(\alpha)},\quad\text{and} \quad\delta\eta_j=\sum_\alpha \delta\eta_\alpha  \bar{\phi}_j^{(\alpha)},
\end{align}
and rewrite Eq.~\eqref{eq:app3Dmodnetyquat} as follows
\begin{align}
\label{eq:app3Dmodnetycin}
\begin{dcases}
 \frac{d\delta{u}_{\alpha}}{dt} = -2\lambda_A \cos(a) \delta u_\alpha -\epsilon \lambda_C \Lambda^{(\alpha)}\left( \cos(c) \delta u_\alpha-\sin(c) \delta v_\alpha\right)\\
 \frac{d\delta{v}_{\alpha}}{dt}  = -2\lambda_A \frac{\cos(a)\sin(b)}{\cos(b)}\delta u_\alpha -\epsilon \lambda_C \Lambda^{(\alpha)}\left( \sin(c) \delta u_\alpha+\cos(c) \delta v_\alpha\right)\\
 \frac{d\delta{\eta}_{\alpha}}{dt}  =\left(\mu_A -\frac{\lambda_A \cos(a)}{\lambda_B \cos(b)}\mu_B\right)\delta\eta_\alpha-\epsilon \mu_C \Lambda^{(\alpha)}\delta\eta_{\alpha}\, .
\end{dcases}
\end{align}

The latter can be written in compact for as
 \begin{eqnarray}
\label{eq:app3Dmodnetysix}
\frac{d}{dt}
\begin{pmatrix}
\delta u_\alpha\\
\delta v_\alpha\\
\delta \eta_\alpha
\end{pmatrix}
 &= &
\begin{pmatrix}
 -2\lambda_A \cos(a) -\epsilon \lambda_C \Lambda^{(\alpha)}\cos(c) & \epsilon \lambda_C \Lambda^{(\alpha)} \sin(c) &0 \\
 -2\lambda_A \frac{\cos(a)\sin(b)}{\cos(b)} -\epsilon \lambda_C \Lambda^{(\alpha)} \sin(c) & -\epsilon \lambda_C \Lambda^{(\alpha)} \cos(c) & 0\\
 0 & 0 & \left(\mu_A -\frac{\lambda_A \cos(a)}{\lambda_B \cos(b)}\mu_B\right)-\epsilon \mu_C \Lambda^{(\alpha)}
\end{pmatrix}
\begin{pmatrix}
\delta u_\alpha\\
\delta v_\alpha\\
\delta \eta_\alpha
\end{pmatrix}\notag\\
&=&\mathbf{J}_\alpha\begin{pmatrix}
\delta u_\alpha\\
\delta v_\alpha\\
\delta \eta_\alpha
\end{pmatrix}\, .
\end{eqnarray}

Let us observe that also in this case, the peculiar shape of the differential equation and the choice of the perturbation, allowed to get a time independent Jacobian matrix $\mathbf{J}_\alpha$ even if resulting from the linearization about a periodic solution. We are thus interested in the eigenvalues of the latter matrix. A straightforward computation allows one to conclude that one eigenvalue is
\begin{align}
 \lambda_1 = \left(\mu_A -\frac{\lambda_A \cos(a)}{\lambda_B \cos(b)}\mu_B\right)-\epsilon \mu_C \Lambda^{(\alpha)}\, ,
\end{align}
and the two other are the solutions of 
\begin{align}
\lambda^2+2\lambda \left(\epsilon \lambda_C \Lambda^{(\alpha)}\cos(c)+\lambda_A\cos(a)\right)+\epsilon^2\lambda_C^2(\Lambda^{(\alpha)})^2+2\epsilon \lambda_A\lambda_C \Lambda^{(\alpha)}\cos(a)\left( \cos(c) +\frac{\sin(b)\sin(c)}{\cos(b)}\right)=0,
\end{align}
namely,

\begin{align}
\label{eq:appsol2ndorder}
 \lambda_{\pm}=-\epsilon \lambda_C \Lambda^{(\alpha)}\cos(c)-\lambda_A\cos(a)\pm \sqrt{[\lambda_A\cos(a)]^2-2\epsilon \lambda_A\lambda_C\Lambda^{(\alpha)}\cos(a)\frac{\sin(b)}{\cos(b)}\sin(c)+[\lambda_C\epsilon \Lambda^{(\alpha)}\cos(c)]^2-[\epsilon \lambda^{(\alpha)}\lambda_C]^2}\, .
\end{align}
Let us observe that once $\Lambda^{(\alpha)}\rightarrow 0^+$, then the eigenvalue $\lambda_+$ vanishes and the other ones are negative if $\epsilon >0$, $\mu_C>0$ and $\Lambda^{(\alpha)}\geq 0$ (recall the stability condition of the solution~\eqref{eq:appcondstab}). The stability of the solution can thus be proved by looking at the sign of $\lambda_+$. One can of course use the explicit formula~\eqref{eq:appsol2ndorder}, there is however another method that we hereby present. By developing $\lambda_+$ for $\Lambda^{(\alpha)}$ close to $0^+$ we get
\begin{align}
 \lambda_+ = -\epsilon \Lambda^{(\alpha)}\lambda_C \left(\cos(c)+\lambda_A\frac{\cos(a)\sin(b)\sin(c)}{\cos(b)}\right)+\mathcal{O}(\Lambda^{(\alpha)})^2\, ,
\end{align}
and, thus, if
\begin{align}
\label{eq:appconditionMSFneg}
\lambda_C \left(\cos(c)+\lambda_A\frac{\cos(a)\sin(b)\sin(c)}{\cos(b)}\right)>0,
\end{align}
then $\lambda_+<0$ for all $\Lambda^{(\alpha)}$ close enough to $0$ (see panel (a) of Fig.~\ref{fig:DispRel}). Let us observe that in the case the latter inequality is not satisfied, then $\lambda_+$ will be positive close to $\Lambda^{(\alpha)}=0$, then the synchronization will depend on the spectrum of the Matrix Weighted Network as well as on $\epsilon$ (see panel (b) of Fig.~\ref{fig:DispRel}).
\begin{figure}[ht!]
    \centering
\subfigure[\hspace{-15pt}]{\includegraphics[width=0.45\textwidth]{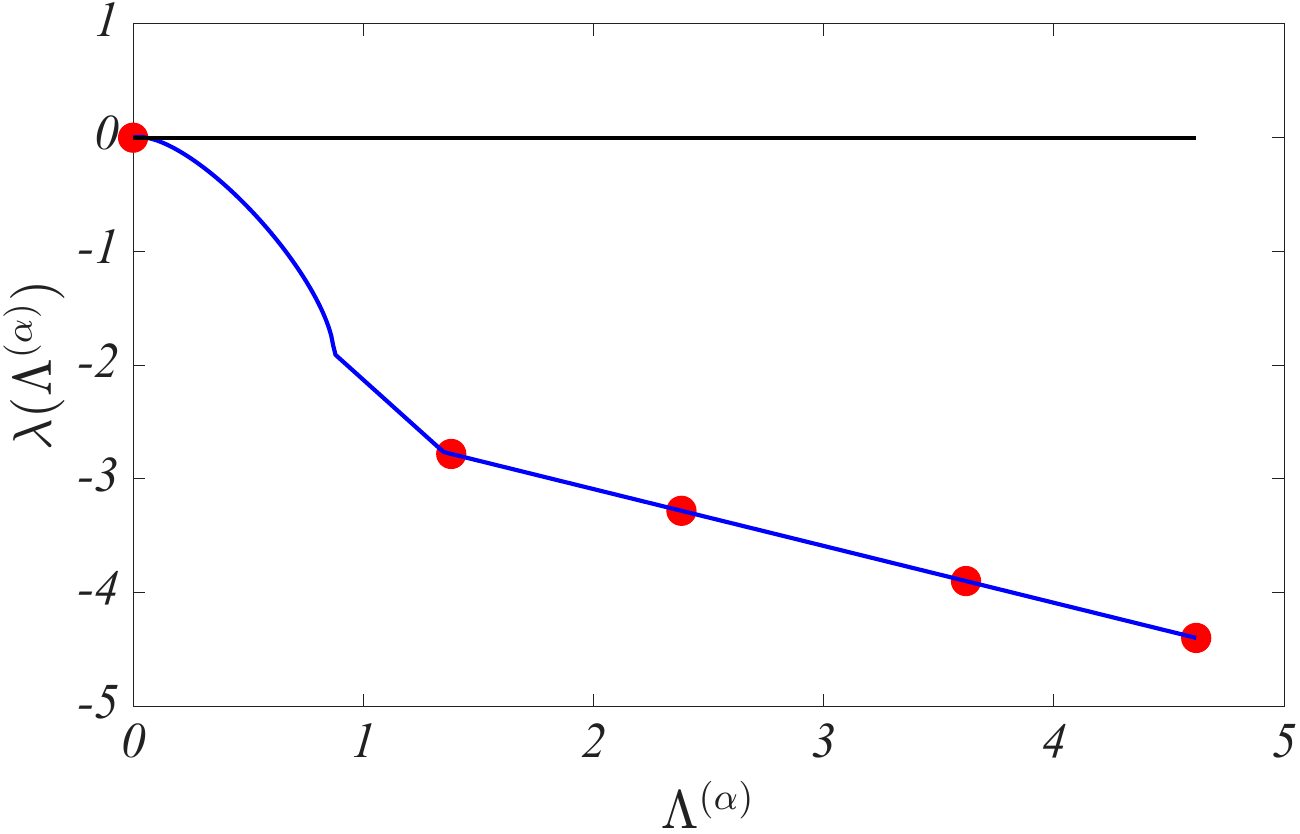}}\quad \subfigure[\hspace{-15pt}]{\includegraphics[width=0.45\textwidth]{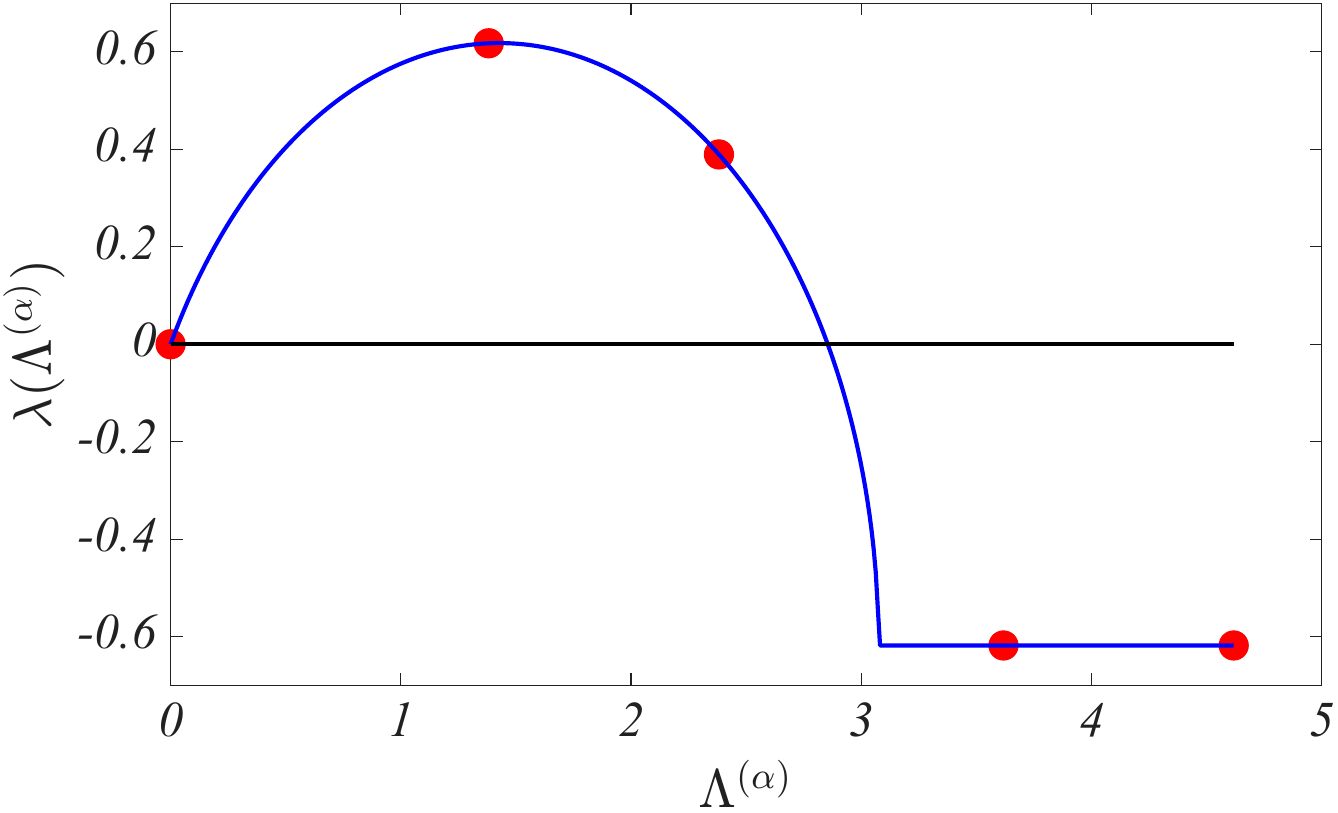}} 
        \caption{Two possible shapes for the Master Stability Function. Parameters have been chosen in such a way condition~\eqref{eq:appconditionMSFneg} is satisfied in panel (a), while it does not on panel (b).}
        \label{fig:DispRel}
\end{figure}

Finally, let us observe that if the matrix $\mathbf{C}$ reduces to a diagonal matrix, $
\begin{pmatrix}
\lambda_C & 0 &0\\
0 & \lambda_C& 0\\
0 & 0 & \mu_C
\end{pmatrix}$, i.e., if $c=0$, then the above condition only relies on the sign of $\lambda_C$.

In Fig.~\ref{fig:NumResEx3Sync}, we report a case where GS emerges for the $\mathrm{MWN_c}$ of Fig. 1c of the Main text for a suitable choice of model parameters, while in Fig.~\ref{fig:NumResEx3NoSync} the opposite case is shown: no global synchronization can emerge.
\begin{figure*}[ht!]
\subfigure[\hspace{-15pt}]{\includegraphics[width=0.3\textwidth]{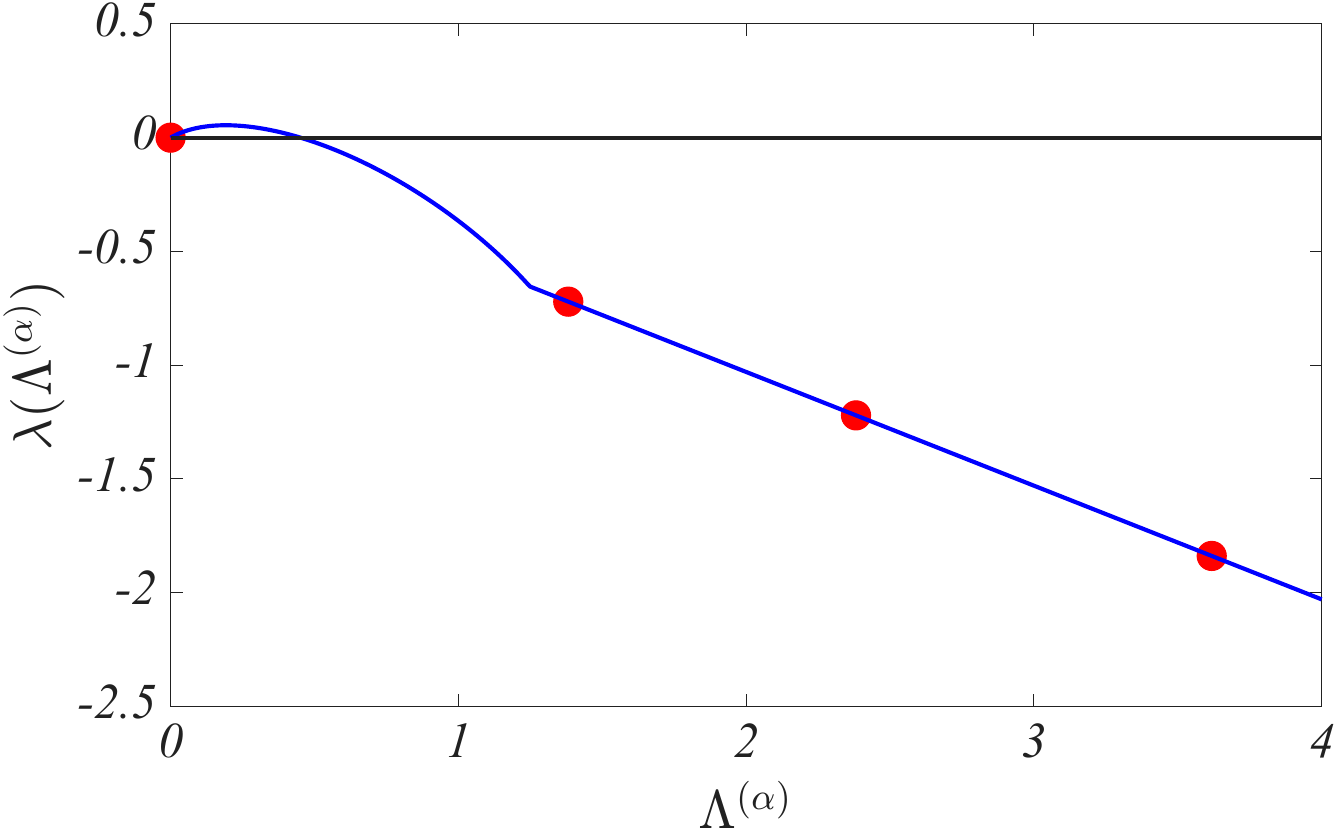}} \hfill        \subfigure[\hspace{-15pt}]{\includegraphics[width=0.33 \textwidth]{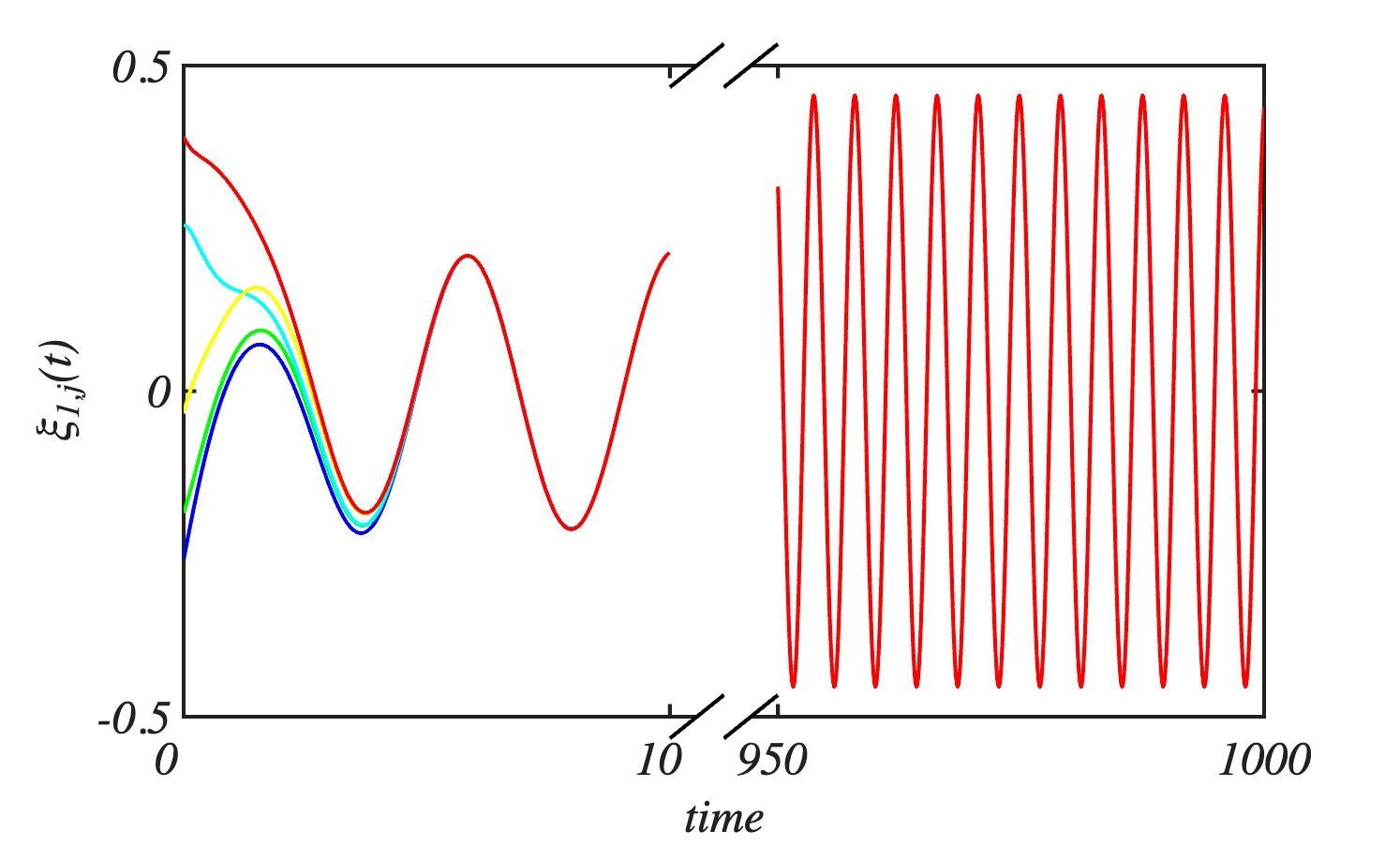}} \hfill \subfigure[\hspace{-15pt}]{\includegraphics[width=0.3\textwidth]{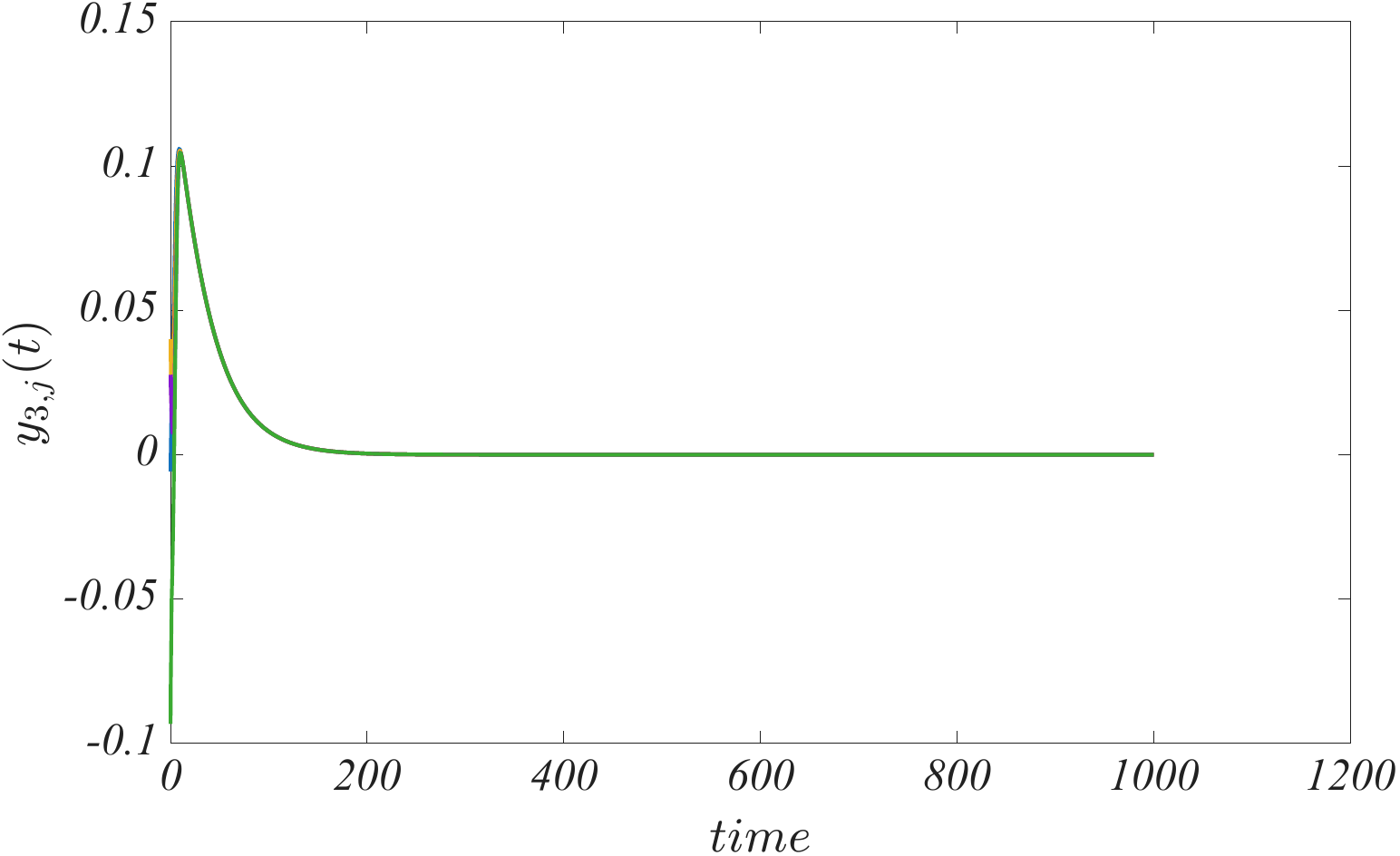}}         \caption{Synchronization of the $3$D-model~\eqref{eq:3Dmodnetyapp} defined on the $\mathrm{MWN_c}$ of Fig. 1c of the Main text. Panel (a): the Master Stability Function, panel (b): $\xi_{1,j}(t)$, i.e., the first component of the vector $\vec{y}_j$ and panel (c): $y_{3,j}(t)$. The $\mathrm{MWN_c}$ has been obtained by setting $\theta=2\pi/3$ and the scalar weights have been set $w_{ij}=1$. The matrices, $\mathbf{A}$, $\mathbf{B}$ and $\mathbf{C}$ are defined with $a = 2\pi/5$, $b = 2\pi/3$, $c = 2\pi/7$, $\lambda_A = 1$, $\lambda_B = -3$, $\lambda_C = 2$, $\mu_A = 1$, $\mu_B = 5$ and $\mu_C = 1$. The coupling parameters is fixed to $\epsilon = 0.5$.}
        \label{fig:NumResEx3Sync}
\end{figure*}

\begin{figure*}[ht!]
    \centering
        \subfigure[\hspace{-15pt}]{\includegraphics[width=0.3\textwidth]{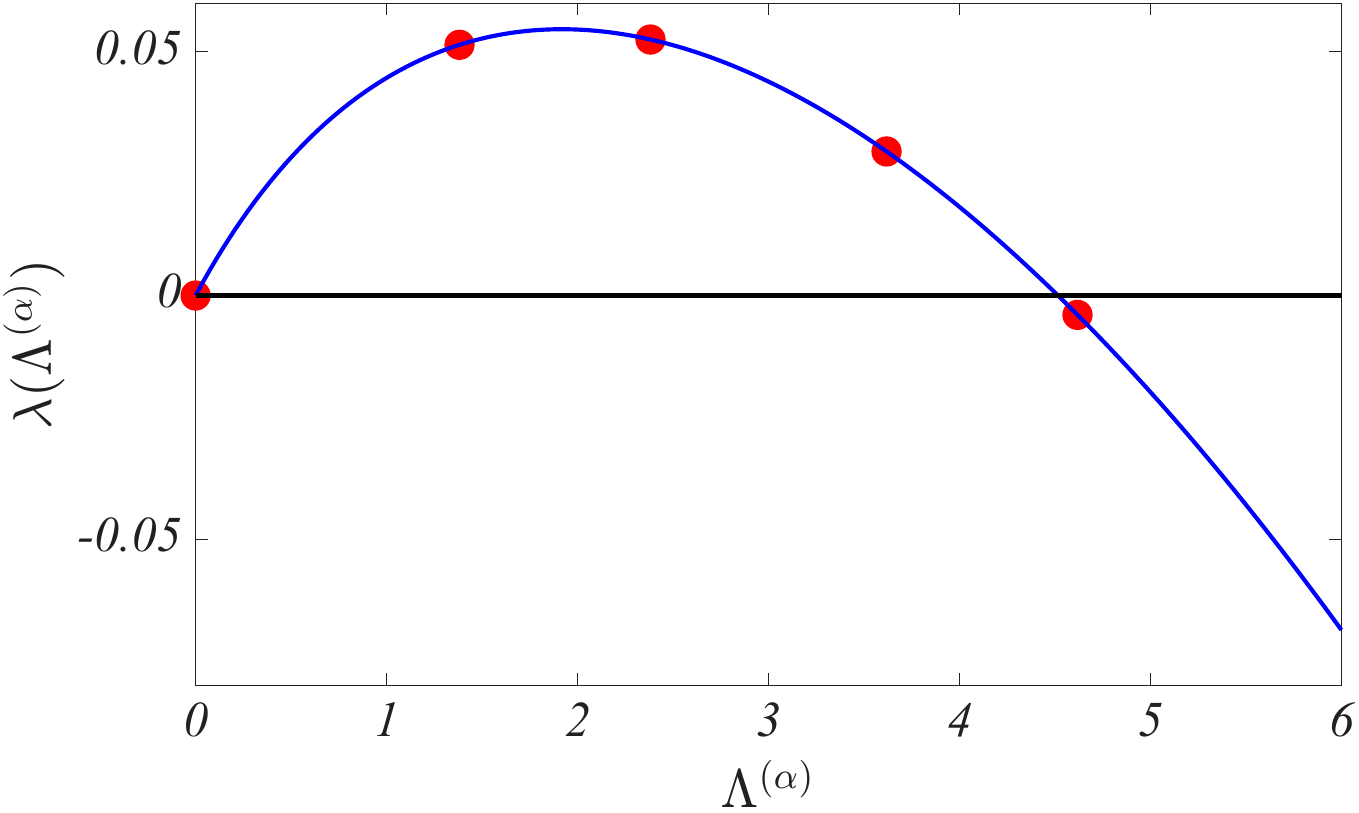}} \quad
        \subfigure[\hspace{-15pt}]{\includegraphics[width=0.33\textwidth]{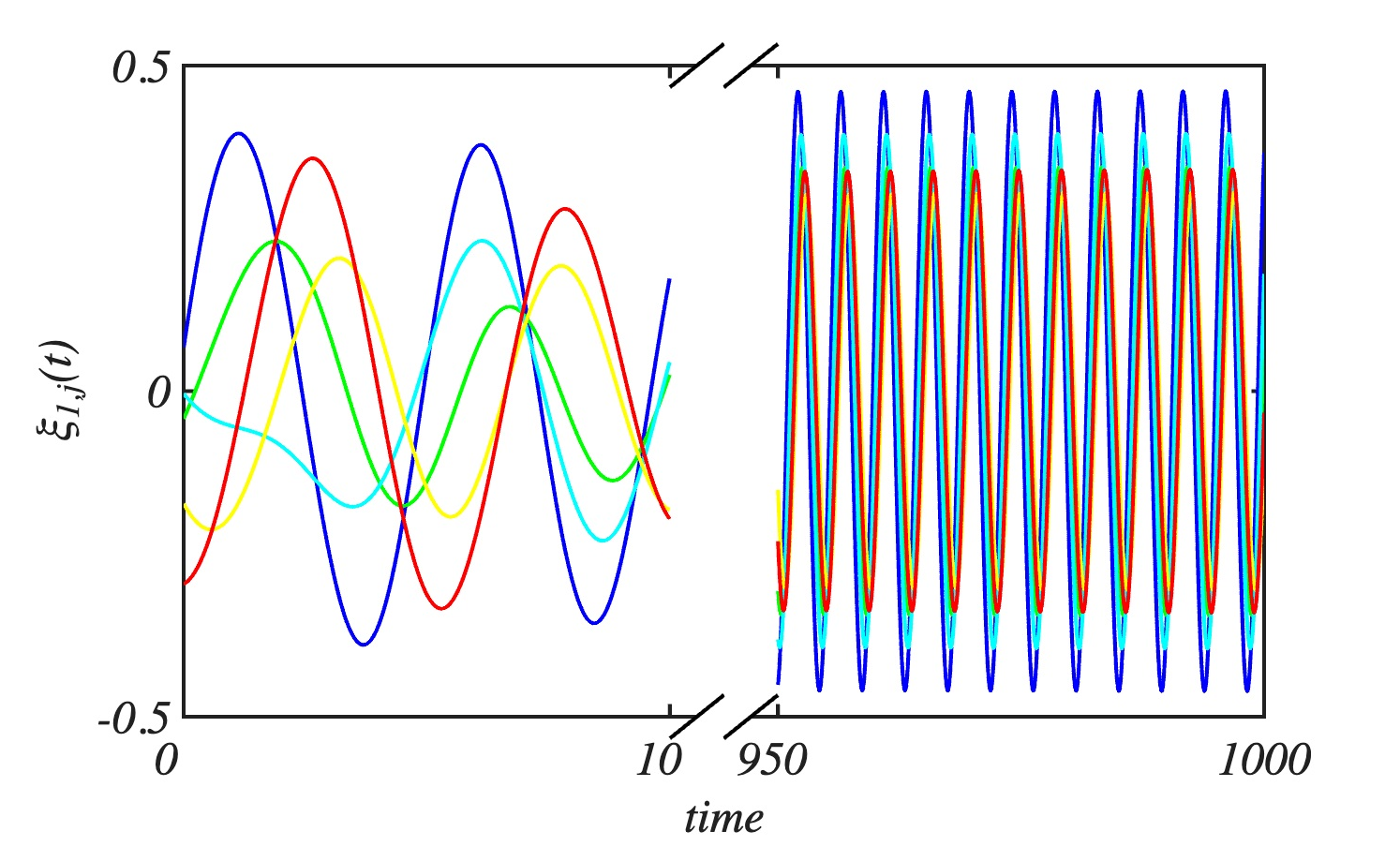}} \quad
        \subfigure[\hspace{-15pt}]{\includegraphics[width=0.3\textwidth]{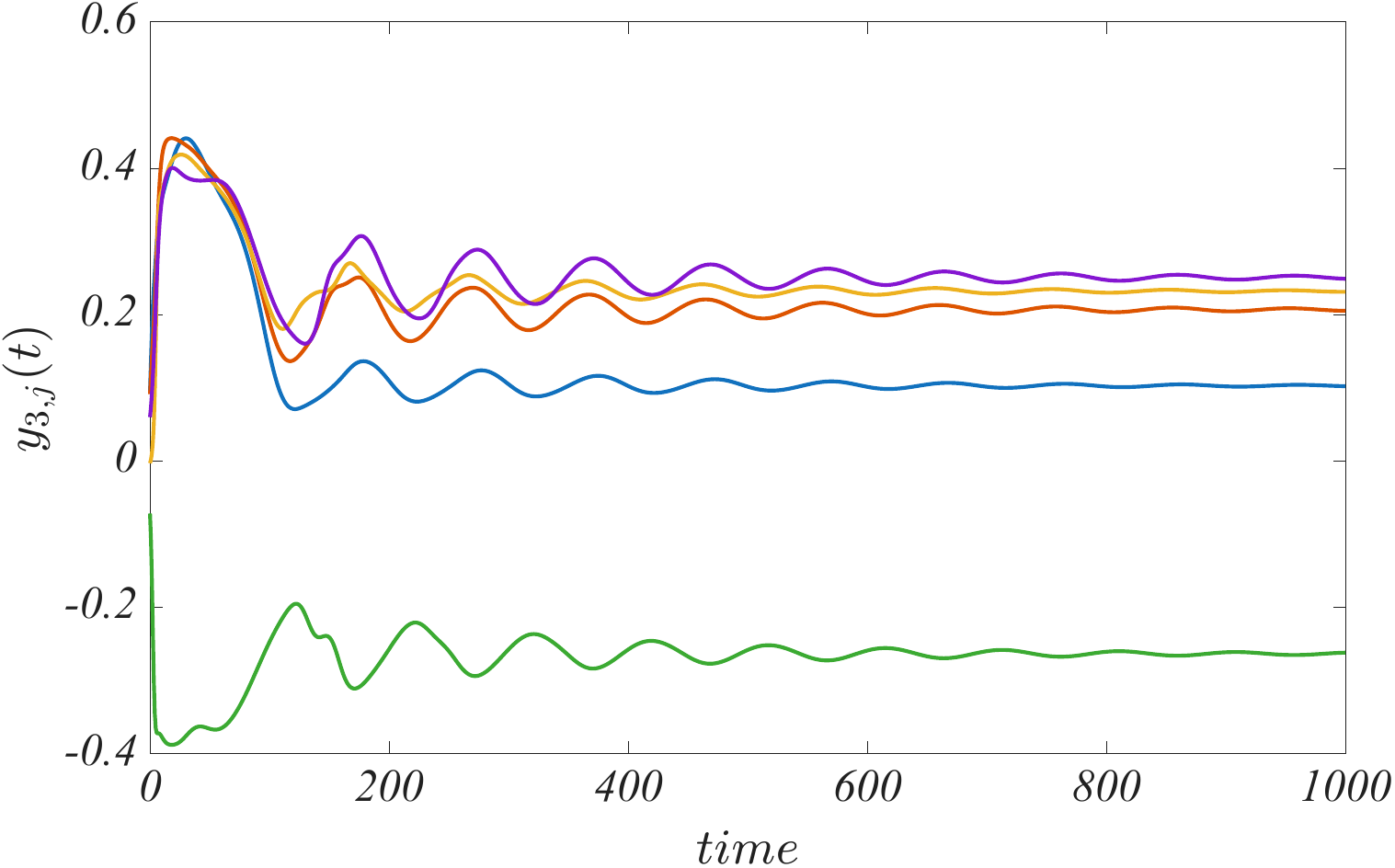}}
        \caption{Non synchronization of the $3$D-model~\eqref{eq:3Dmodnetyapp} defined on the $\mathrm{MWN_c}$ of Fig. 1c of the Main text. Panel (a): the Master Stability Function, panel (b): $\xi_{1,j}(t)$, i.e., the first component of the vector $\vec{y}_j$, and panel (c): $y_{3,j}(t)$. The $\mathrm{MWN_c}$ has been obtained by setting $\theta=2\pi/3$, and the scalar weights have been set $w_{ij}=1$. The matrices, $\mathbf{A}$, $\mathbf{B}$ and $\mathbf{C}$ are defined with $a = 2\pi/5$, $b = 2\pi/3$, $c = 2\pi/7$, $\lambda_A = 1.5$, $\lambda_B = -3$, $\lambda_C = 2$, $\mu_A = 1$, $\mu_B = 5$ and $\mu_C = 1$. The coupling parameters is fixed to $\epsilon = 0.05$.}
        \label{fig:NumResEx3NoSync}
\end{figure*}

\begin{remark}\label{rem:generalaxis} 
    \emph{For any $\vec v\in\mathbb R^3$, there exists transformation $\mathbf Q\in SO(3)$, that can be constructed by choosing an orthonormal basis $\{v_1,v_2,v_3\}\subset\mathbb R^3$, with $v_3\equiv\vec u$, and defining $\mathbf Q$ as the matrix whose rows are $v_1^\top,v_2^\top,v_3^\top$, such that $\mathbf Q\vec v=\vec u$ where $\vec u=(0,0,1)^\top$. This implies that a rotation about an arbitrary axis $\vec v\in\mathbb R^3$ can be transformed into a rotation about $\vec u$ via a similarity transformation.} 

    \emph{Indeed, for any $\vec w\in\mathbb R^3$, $[\vec v]_\times\vec w\equiv\vec v\times\vec w$ and thus, exploiting the fact that rotation matrices preserve the cross product, we have}

    \begin{align}\label{eq:skew1}
        [\vec u]_\times\vec w=\mathbf Q[\vec v]_\times\vec w=\mathbf Q(\vec v\times\vec w)=(\mathbf Q\vec v)\times(\mathbf Q\vec w)=\vec u\times(\mathbf Q\vec w)=[\vec u]_\times \mathbf Q\vec w,
    \end{align}
    \emph{i.e., the skew-matrix $[\vec v]_\times$ satisfies the identity $\mathbf Q[\vec v]_\times\mathbf Q^\top=[\vec u]_\times$. Furthermore, it holds} 

    \begin{align}\label{eq:skew2}
        (\mathbf Q[\vec v]_\times \mathbf Q^\top)^2=\mathbf Q[\vec v]_\times\mathbf Q\mathbf Q^\top[\vec v]_\times\mathbf Q^\top= \mathbf Q[\vec v]_\times^2\mathbf Q^\top.
    \end{align}
    \emph{From Eqs.~\eqref{eq:skew1} and~\eqref{eq:skew2}, it follows that}

    \begin{align}
        \mathbf Q\mathbf R(\vec v,\theta)\mathbf Q^\top
        = I+\sin(\theta)\mathbf Q[\vec v]_\times\mathbf Q^\top+(1-\cos(\theta))Q[\vec v]_\times^2 \mathbf Q^\top
        =I+\sin(\theta)[\vec u]_\times+(1-\cos(\theta))[\vec u]_\times^2 = \mathbf R(\vec u,\theta),
    \end{align}
    \emph{for any $\theta\in[0,2\pi)$. Therefore, for any $\vec v\in\mathbb R^3$, one can always reduce the problem to the case of a rotation about the $\vec u$-axis.}
\end{remark}

\section*{Supplementary Note 3:\\ Another example of linear couplings for the Lorenz model defined on MWN}
\label{sec:appmoreLorenz}

The aim of this section is to provide another example of linear coupling for the Lorenz model defined on MWN and to show that global synchronization can be achieved. For the sake of simplicity we used the same MWN presented in Section III.C in the case of Lorenz system, i.e., a square with rotation matrices performing the transformation $(x,y,z)^\top\rightarrow(-x,-y,z)^\top$.

In panel (a) of Fig.~\ref{fig:LorenzSynch2} we report the MSF as a function of $\kappa = \epsilon \Lambda^{(\alpha)}$, in the case of the $1-2$ coupling, namely $E_{12}=1$ and all the remaining entries vanish. We can observe that the MSF is positive for $\kappa=0$, testifying the chaotic behavior of the reference solution $\vec{s}(t)$, then it decreases, vanishes at $\kappa_1\sim 4.098$, reaches a minimum and then increases. A second zero can be found for $\kappa_2\sim 22.640$. This implies that global synchronization can emerge if and only if $2\epsilon > \kappa_1$ and $4\epsilon < k_2$, remember that the eigenvalues of the Laplace matrix associated to the MWN are given by $\{0,2,4\}$. The case reported in panel (a) of Fig.~\ref{fig:LorenzSynch2} corresponds to $\epsilon = 5$, that satisfies the above constraints (see red circles). The system achieves thus global synchronization as one can appreciate from panel (b) where we report the first component of $\vec{y}_j=\mathbf{R}_\pi \vec{x}_j$ as a function of time, and also from panel (c) depicting the synchronization error in the original variables (blue curve) and in the rotated ones (red curve). The very small values assumed in the latter case, confirm the global synchronization.

\begin{figure*}[ht!]
\includegraphics[width=0.9\textwidth]{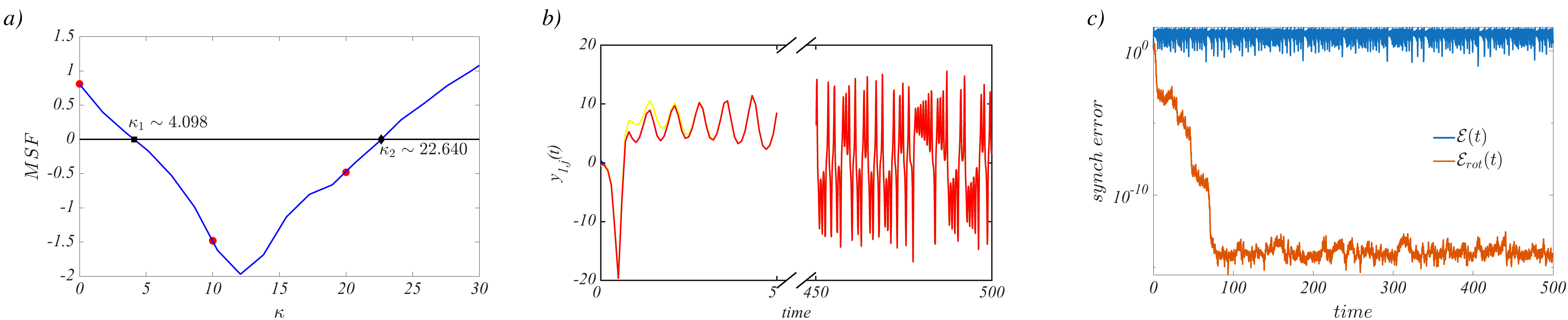}      
\caption{\textbf{Global synchronization of Lorenz oscillators $2-1$ coupled with a square MWN}. 
In panel (a) we report the Master Stability Function, in panel (b) the ${y}_{1,j}(t)$, i.e., the first component of $\vec{y}_j$, while in panel (c) we report the time evolution of the synchronization error computed in the original $\vec{x}_i$ variables (blue curve) and in the rotated ones, $\vec{y}_i$ (red curve). The coupling parameter is fixed to $\epsilon = 5$.
}
\label{fig:LorenzSynch2}
\end{figure*}

We conclude this section by showing once again that coherence is a necessary condition for the emergence of global synchronization also in the case of chaotic oscillators. We thus consider again the Lorenz model presented in the Section III.C in the main text, but we slightly perturb the rotation matrix (see panel (a) of Fig.~\ref{fig:LorenzSynch3}) in such a way the MWN is no longer coherent. Without changing any other parameters, global synchronization cannot now be achieved as one can observe from the large values of the synchronization errors (blue and red curves in panel (c) of Fig.~\ref{fig:LorenzSynch3}).
\begin{figure*}[ht!]
\includegraphics[width=0.9\textwidth]{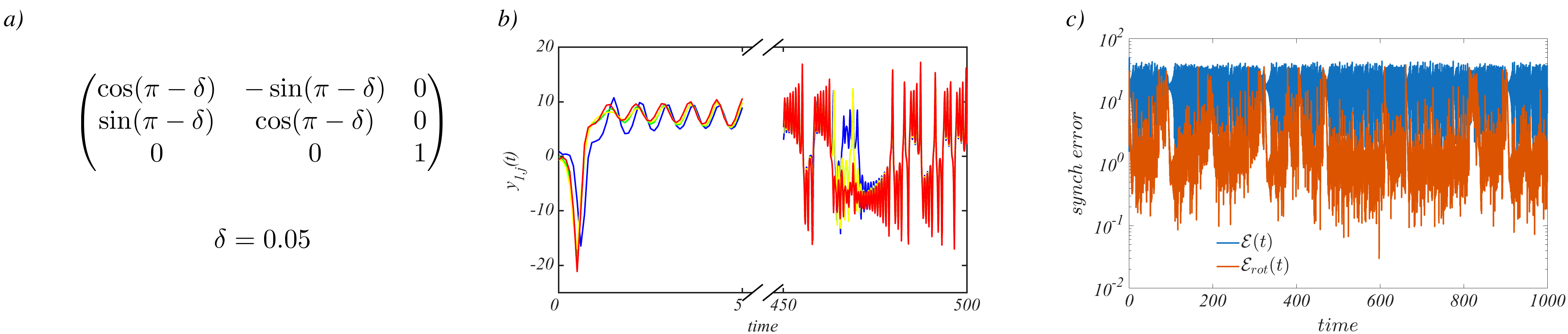}      
\caption{\textbf{Absence of Global synchronization of Lorenz oscillators $1-1$ coupled with a square MWN}. 
In panel (a) we report the rotation matrix, in panel (b) the ${y}_{1,j}(t)$, i.e., the first component of $\vec{y}_j$, while in panel (c) we report the time evolution of the synchronization error computed in the original $\vec{x}_i$ variables (blue curve) and in the rotated ones, $\vec{y}_i$ (red curve). The coupling parameter is fixed to $\epsilon = 3$.
}
\label{fig:LorenzSynch3}
\end{figure*}

\end{document}